\documentclass[apl,showpacs,superscriptaddress,amsmath,amssymb,twocolumn]{revtex4-1}

\usepackage{graphicx}
\usepackage{epsfig}
\usepackage{multirow}
\usepackage{array}
\usepackage{booktabs}
\usepackage{float}
\usepackage{subfigure} 

\begin{document}

\title{A study of size-dependent properties of MoS$_{2}$ monolayer nanoflakes using density-functional theory}

\author{M. Javaid}
\email{maria.javaid@rmit.edu.au}
\affiliation{Chemical and Quantum Physics, School of Science, RMIT University, Melbourne VIC 3001, Australia}
\affiliation{The Australian Research Council Centre of Excellence for Nanoscale BioPhotonics, School of Science, RMIT University, Melbourne, VIC 3001, Australia}

\author{Daniel W. Drumm}
\affiliation{The Australian Research Council Centre of Excellence for Nanoscale BioPhotonics, School of Science, RMIT University, Melbourne, VIC 3001, Australia}
\author{Salvy P. Russo}
\affiliation{Chemical and Quantum Physics, School of Science, RMIT University, Melbourne VIC 3001, Australia}
\affiliation{ARC Centre of Excellence in Exciton Science, School of Science, RMIT University, Melbourne, VIC 3001, Australia}

\author{Andrew D. Greentree}
\affiliation{Chemical and Quantum Physics, School of Science, RMIT University, Melbourne VIC 3001, Australia}
\affiliation{The Australian Research Council Centre of Excellence for Nanoscale BioPhotonics, School of Science, RMIT University, Melbourne, VIC 3001, Australia}

\date{\today}

\begin{abstract}
Novel physical phenomena emerge in ultra-small sized nanomaterials. We study the limiting small-size-dependent properties of MoS$_{2}$ monolayer rhombic nanoflakes using density-functional theory on structures of size up to Mo$_{35}$S$_{70}$ (1.74~nm). We investigate the structural and electronic properties as functions of the lateral size of the nanoflakes, finding zigzag is the most stable edge configuration, and that increasing size is accompanied by greater stability. We also investigate passivation of the structures to explore realistic settings, finding increased HOMO-LUMO gaps and energetic stability. Understanding the size-dependent properties will inform efforts to engineer electronic structures at the nano-scale.
\end{abstract}
\maketitle
\section{Introduction}
Recently two-dimensional (2D) materials have drawn significant interest due to their unique  structural, electronic, and optical properties \cite{Novoselove,atlas2D,TMDCs_photonics}. The existence of 2D materials had been a highly debated issue until the successful exfoliation of graphene from graphite, the first experimentally stable 2D material \cite{Geim}. After this revolutionary discovery, many other 2D materials such as silicene, hexagonal boron nitride and transition-metal dichalcogenides (TMDCs) have also been exfoliated \cite{Li201533}. These 2D materials are now a widely growing field with a diverse range of applications in nano-electronics \cite{TMDCs_photonics}. 

TMDCs belong to a family of layered materials where each layer is connected through weak Van der Waals forces. They have a general formula of MX$_{2}$, where M is a transition metal (M = Mo, W, Zr, Hf, \textit{etc.}) and X is a chalcogen (X = S, Se, Te, \textit{etc.}). Each layer is three atoms thick with the metal in the centre and the chalcogen atoms above and below the metal \cite{GeneralFormula}. Nanoflakes of these materials are promising due to the properties emerging from their inter-layer or intra-layer bonding \cite{NanoflakesMiro}. Property variations emerge by changing the number of layers or the lateral size within a layer. For example, bulk MoS$_{2}$ has an indirect band gap of 1.2 eV but when it is thinned down to a single layer, its band gap switches to a direct band gap of 1.88 eV which makes it promising for photonic applications \cite{2010Nano10,PhysRevLett}. However, description of the consequences of lateral size variation in small sized MoS$_{2}$ monolayer flakes are as yet incomplete.

MoS$_{2}$ is a compound which belongs to the hexagonal $P6_{3}/mmc$ space group. In its layered structure, each S atom is covalently bonded to three Mo atoms and each Mo atom to six S atoms forming a trigonal prismatic coordination \cite{Kadantsev2012909}. The symmetry group of monolayer MoS$_{2}$ is $D_{3h}^{1}$ which contains the discrete symmetries: $C_{3}$ trigonal rotation, $\sigma_{h}$ reflection by the $xy$ plane, $\sigma_{v}$ reflection by the $yz$ plane, and all of their products \cite{XLF2012}. 

\begin{figure}[tb]
\centering
\includegraphics[scale=0.5, height= 6 cm]{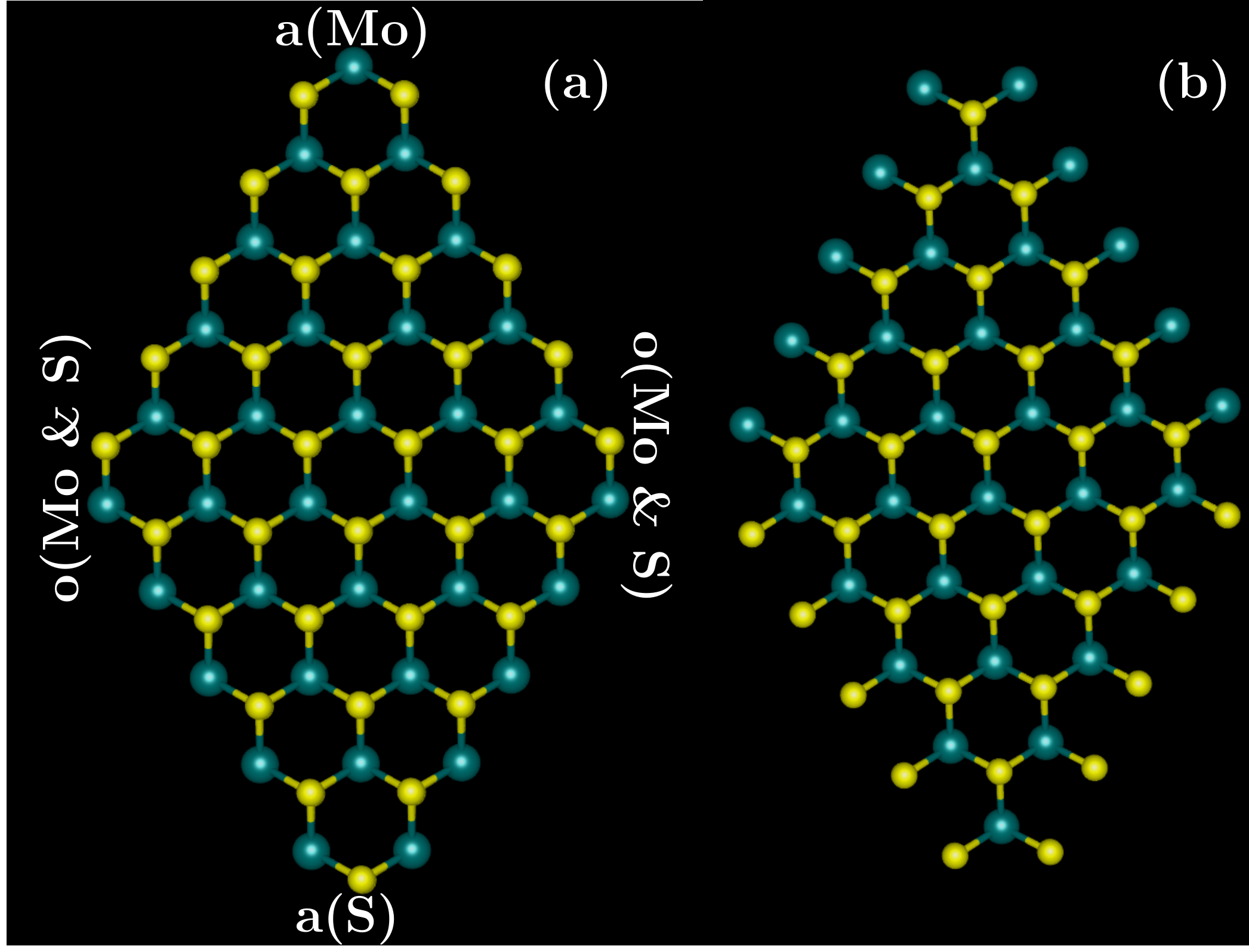}
\caption{Nanoflakes of MoS$_{2}$ monolayer having 105 atoms before geometry optimization: (a) zigzag edge configuration; (b) armchair edge configuration. Large, green atoms are Mo and small, yellow are S. Corner labels are defined as; a(Mo) = acute-Mo, a(S) = acute-S, o(Mo \& S) = obtuse-Mo and S.}
\label{fig:Edges}
\end{figure}

There have been significant efforts to understand the size- and edge-dependent,  structural and electronic properties of MoS$_{2}$ monolayer nanoflakes. For example, quantum confinement effects in TMDC nanoflakes have been investigated by Mir\'{o} \textit{et al.},~both experimentally and through density-functional theory (DFT) \cite{NanoflakesMiro}. Wendumu \textit{et al.}~have presented the size-dependent optical properties of 1.6 to 10.4 nm MoS$_{2}$ nanoflakes \cite{Wendumu2014} using the density-functional tight-binding (DFTB) method. An extensive DFT edge-dependence study on MoS$_{2}$ monolayer nanoribbons has been reported by Pan \textit{et al.}~\cite{PZ2012}. Recently Nguyen \textit{et al.}~have experimentally studied the size-dependent properties of few-layer MoS$_{2}$ nanosheets and nanodots \cite{Nguyen2016} but a complete study of the structural, electronic and optical properties of very small single-layer MoS$_{2}$ nanoflakes has not yet been presented.
\begin{table}[tb]
\newcolumntype{P}[1]{>{\centering\arraybackslash}p{#1}}
\newcolumntype{M}[1]{>{\centering\arraybackslash}m{#1}}
\centering
\caption{Mean displacement, $\Delta R$, of atoms in the central zone of an optimized 72-atom flake (shown by the red-dashed circle in Fig.~\ref{fig:core_shell}(d)) from the bulk experimental positions of MoS$_{2}$ using several functionals in \textsc{gaussian09}. All functionals except B3LYP predict mean displacements less than 5\% from the bulk values.}
\begin{tabular}{ |M{2.5cm}|M{2.5cm}| }
 \toprule
 Functionals  & $\Delta R$\\
 \midrule
\hline
 
PBE1PBE & 0.0256\\ 
 B3LYP&  0.0565\\
 BHandHLYP & 0.0400\\
M052X   &0.0330\\
 \bottomrule
\end{tabular}
\label{table:Tab1}
\end{table}

Here we report a DFT study of the 0 K size-dependent properties of 1H MoS$_{2}$ monolayers of size smaller than 2~nm. Although we have exclusively studied MoS$_{2}$ structures, our approach can be generalized to other TMDC nanoflakes of similar size. We begin our discussion by studying the relative stability of the armchair and zigzag configurations. We present the geometries of the relaxed structures for different nanoflake sizes to thoroughly understand the structural response as a function of lateral size. We report the electronic properties; binding energy, flake formation energy, HOMO-LUMO (highest-occupied molecular orbital to lowest-unoccupied molecular orbital) gap, charge densities, and the passivation of the flakes. 

This paper is organized as follows: first we discuss all the required methods and techniques. Then we study two different edge configurations for MoS$_{2}$ monolayers and find the most stable one, following with the discussion of structural stability as a function of size, the electronic properties and the  properties of the passivated structures.
  
\section{Methodology}
\begin{figure}[h!]
\centering
\includegraphics[scale=0.7]{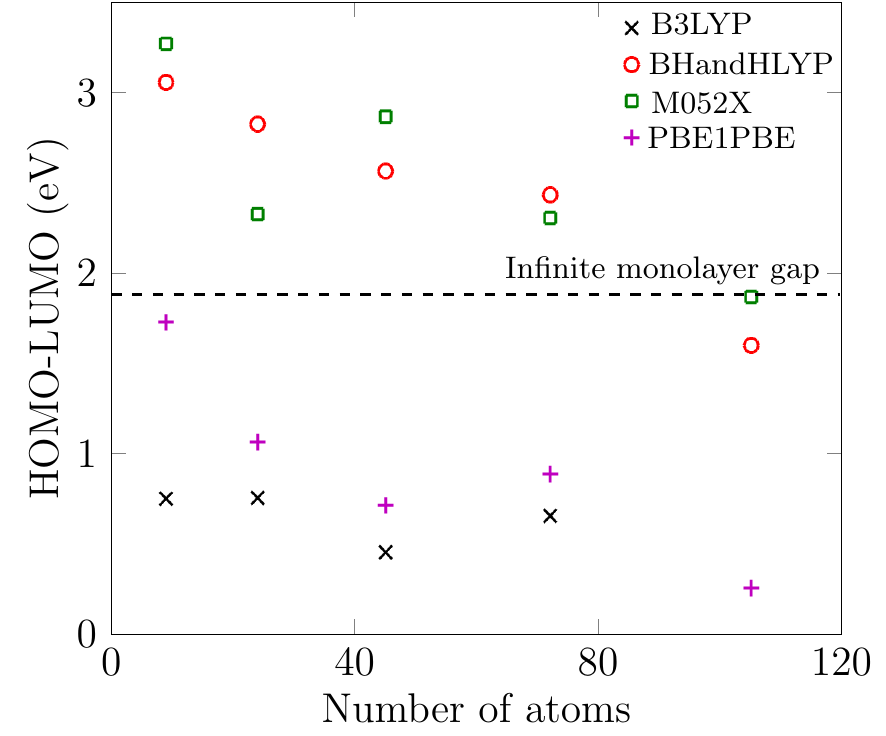}
\caption{A size-dependent analysis of the HOMO-LUMO gap in MoS$_{2}$ monolayer nanoflakes using four different functionals. The black-dashed line is the known experimental gap in an infinitely large sheet of the MoS$_{2}$ monolayer \cite{PhysRevLett}.}
\label{Fig:Functional_HL}
\end{figure}
We investigated the structural and electronic properties of neutral MoS$_{2}$ monolayer nanoflakes with stoichiometry Mo$_{n}$S$_{2n}$ using DFT in \textsc{gaussian09} \cite{gaussiao09}. In experiments, usually triangular shaped islands of MoS$_{2}$ have been reported but it has been theoretically speculated that MoS$_{2}$ islands can exist in various shapes, such as trigonal, hexagonal, truncated hexagonal and rhombohedral \cite{Bertram2006,Lauritsen2007,Halveg2000,Seifert2006}. We used rhombic flakes to maintain the neutrality and Mo$_{n}$S$_{2n}$ stoichiometry of the flakes. Also, we experienced convergence issues with the triangular shaped flakes.

\begin{figure*}[tb!]
\centering
\includegraphics[scale=0.7]{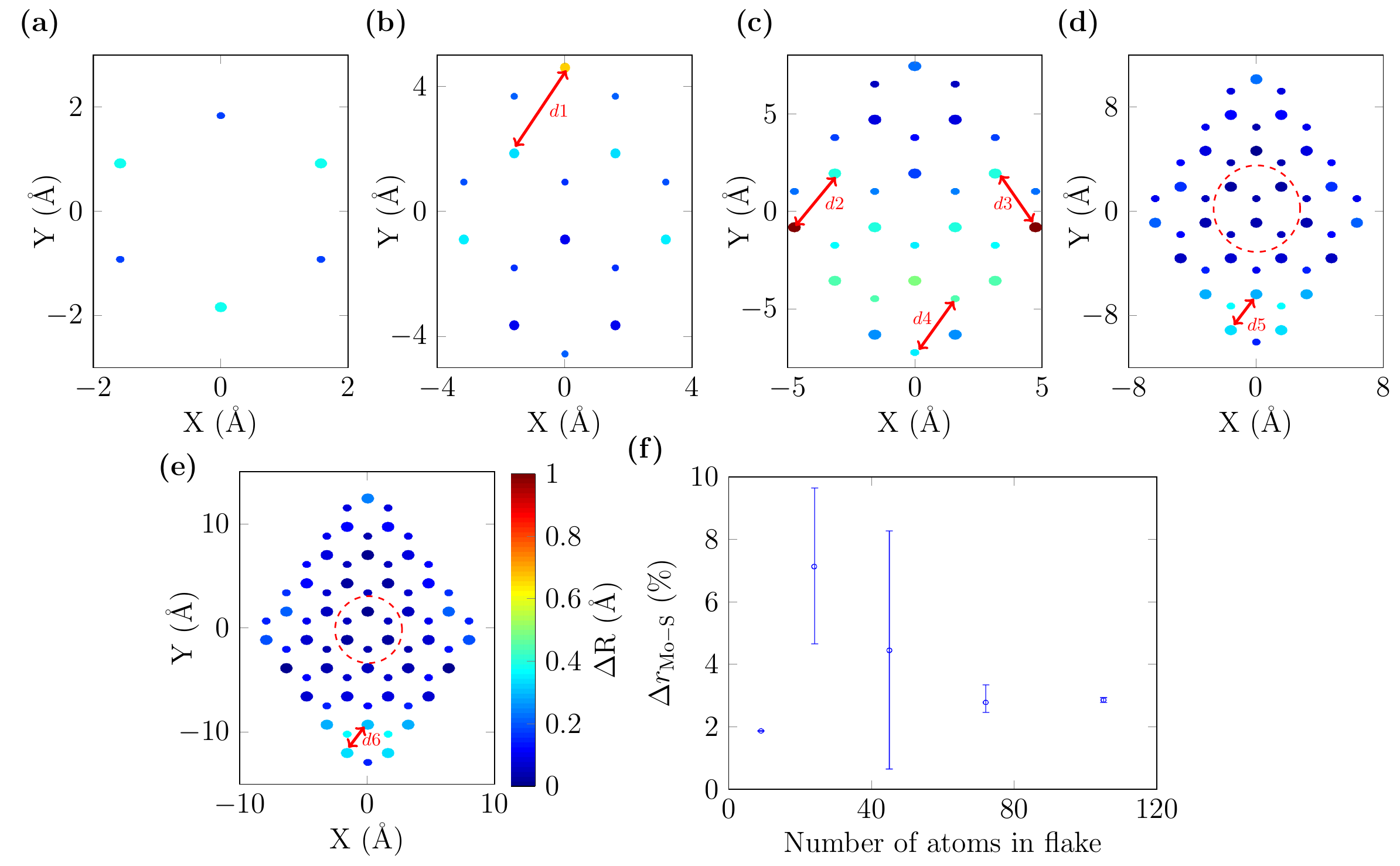}
\caption{Relaxed structures of MoS$_{2}$  monolayer nanoflakes comprised of: (a) 9 atoms, (b) 24 atoms, (c) 45 atoms, (d) 72 atoms, and (e) 105 atoms. The larger circles are Mo and the smaller are S. The colour of the atoms ($\Delta R$ given by Eq.~\ref{Eq:Colour}) represents variation of the atomic positions of relaxed structures from the bulk experimental positions \cite{Dickinson1923}. S atoms are on top of each other along \textit{z}-axis. Colour bar in (e) and labels in Fig.~\ref{fig:Edges}(a) apply to all subfigures (a-e). The most distorted lengths in each flake are shown by the red-arrowed lines, $d1-d6$. (f) Percentage variation of the mean Mo--S bond length in the central zone of each flake from the bulk value \cite{Dickinson1923}. Error bars are extended to the minimum and maximum Mo--S bond lengths in each central zone. Central zone for (a) and (c) is defined similar to that encircled red-dashed in (e), while for (b), it is similar to encircled red-dashed in (d).}
\label{fig:core_shell}
\end{figure*}

To choose an appropriate functional for our modelling, we conducted an in-depth analysis of the functionals listed in Table \ref{table:Tab1}. We picked a relaxed 72-atom flake as this was the largest size we could model with the B3LYP functional. We compared the relative atomic positions of each atom in the central zone of the 72-atom flake (encircled by red dashes in Fig.~\ref{fig:core_shell}(d)) with the bulk structure (infinitely large and regular structure in all three dimensions) \cite{Dickinson1923}. The displacement $\Delta R_{i}$ of each atom from the bulk position is defined as

\begin{equation}
\begin{split}
\Delta R_{i} = \\
\sqrt{(X_{\text{opt}_{i}} -X_{\text{bulk}})^2+(Y_{\text{opt}_{i}} -Y_{\text{bulk}})^2+(Z_{\text{opt}_{i}} -Z_{\text{bulk}})^2},
\label{Eq:Colour}
\end{split}
\end{equation}

where $i$ indexes the atoms in the central zone of the 72-atom flake. The mean value of $\Delta R_{i}$, \textit{i.e.},~$\Delta R$ for each functional is given in Table \ref{table:Tab1}. All functionals except B3LYP \cite{Becke88, ECLYP, B3LYP} result in less than 5\% variation from the bulk atomic positions. This indicates that the three functionals, BHandHLYP \cite{Functional}, PBE1PBE \cite{PBE1PBE}, and M052X \cite{M052XFunctional} predict similar structures at similar levels of accuracy.

We also calculated the HOMO-LUMO gap as function of flake size for all these functionals as shown in Fig.~\ref{Fig:Functional_HL}. We expect the HOMO-LUMO gap to decrease with increasing flake size, approaching the infinite monolayer MoS$_{2}$ gap for larger flakes as reported by Gan \textit{et al.} \cite{Gan2015} through an analytical equation for MoS$_{2}$ monolayer quantum dots of size from 2 nm to 10 nm. Although our flakes are smaller than 2~nm and we are modelling in DFT, nevertheless we expect a similar trend of approximately decreasing bandgap with increasing flake size. Due to the different methods involved, we only compare the trends, not the absolute values of the HOMO-LUMO gaps. B3LYP and PBE1PBE produce HOMO-LUMO gaps well below the known experimental gap for an infinitely large MoS$_{2}$ monolayer (Fig.~\ref{Fig:Functional_HL}). Hence, we do not consider these two functionals further. For smaller flakes, BHandHLYP and M052X both produce HOMO-LUMO gaps well above the infinite monolayer experimental value \cite{PhysRevLett} and we can expect the band gap with these functionals to converge close to the infinite monolayer band gap for larger flakes. It has been reported that M052X is not a recommended functional for transition metal chemistry \cite{M052XReview}. Considering this, we therefore used the BHandHLYP functional for this article, although we have also performed all the calculations with M052X functional and did not find any major difference in the results. A table on the HOMO-LUMO responses of the smallest MoS$_{2}$ monolayer nanoflake for several functionals (in the Appendix) also provided us with guidance for the optimal choice of functional for our DFT modelling.

The hybrid DFT functional, BHandHLYP \cite{Functional}, includes a mixture of Hartree-Fock exchange with the DFT exchange-correlation via the relation 
\begin{equation}
 \text{BHandHLYP}: 0.5E_{x}^{\rm HF} +0.5E_{x}^{\rm LSDA}+
 0.5\Delta E_{x}^{\rm Becke88}+ E_{c}^{\rm LYP};
\end{equation}
$E_{x}^{\rm HF}$ is the Hartree-Fock exchange term, $E_{x}^{\rm LSDA}$ is the Slater local exchange term \cite{Slater}, $\Delta E_{x}^{\rm Becke88}$ is Becke's 1988 \cite{Becke88} gradient correction to the local-spin density approximation (LSDA) for the exchange term, and $E_{c}^{\rm LYP}$ is the Lee-Yang-Parr correlation term \cite{ECLYP}.

\begin{figure}[tb]
\centering
\includegraphics[scale=0.85]{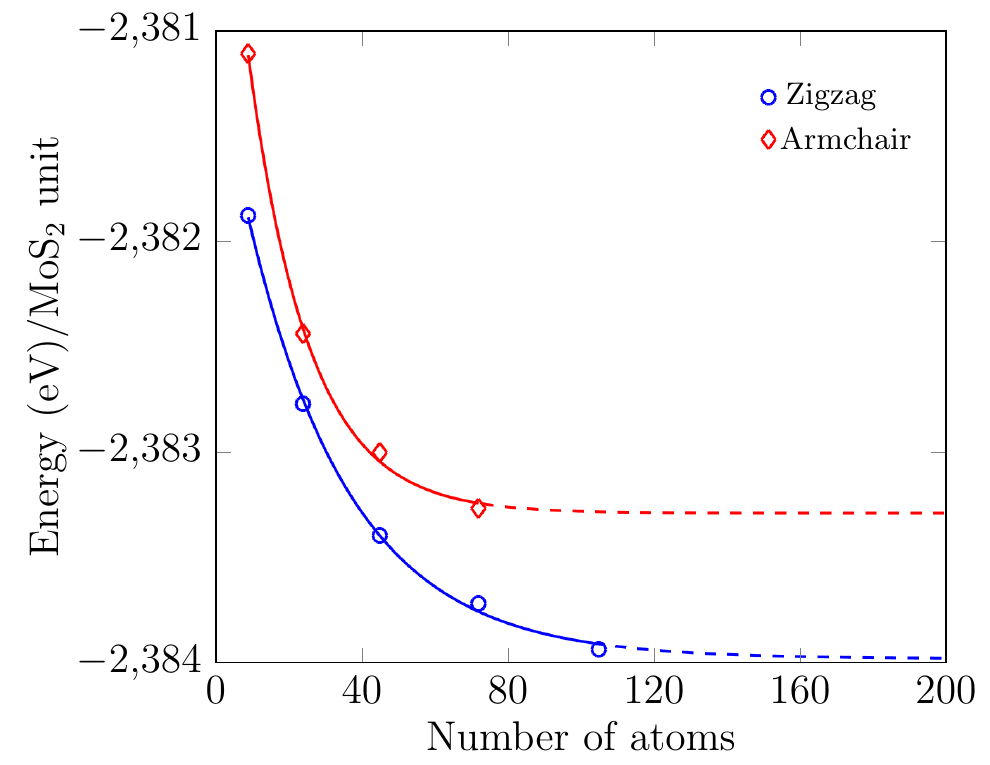}
\caption{Ground-state energies as functions of size. Blue circles represent the zigzag edge configuration and red diamonds the armchair configuration. The solid lines are exponential fits to the data and the dashed lines are extrapolations of those fits for larger structures.}
\label{fig:energetics}
\end{figure}

The basis set used was an effective-core potential basis set of double-zeta quality, the Los Alamos National Laboratory basis set also known as LANL2DZ \cite{LANL2DZ} and developed by Hay and Wadt \cite{HAY85,HAY85a,Wadt85}. These basis sets are widely used in the study of quantum chemistry, particularly for heavy elements \cite{LANL2DZ}. 

\textsc{gaussian09} optimization criteria: calculations were converged to less than 4.5$\times10^{-3}$~Hartree/Bohr maximum force, 3$\times10^{-4}$~Hartree/Bohr RMS force, 1.8$\times 10^{-3}$~Hartree maximum displacement and 1.2$\times10^{-3}$~Hartree RMS displacement. All the flakes were converged to the default SCF (self-consistent field) limit of $<$ 10$^{-8}$ RMS change in the density matrix except those specified in the next section. The charge multiplicity (net charge) was 0 and the spin multiplicity was 1 (singlet; spin neutral).  

In the geometry optimization process, the geometry was modified until a stationary point on the potential surface was found. Analytic gradients were used and the optimization algorithm was the Berny algorithm using GEDIIS \cite{BernyGEDIIS}. We calculated the electronic properties of the optimized structures. The charge densities were plotted in \textsc{avogadro} \cite{Avogadro, Hanwell2012} from a compatible \textsc{gaussian09} checkpoint file. 

\section{Size-dependent structural properties}

The properties of MoS$_{2}$ monolayers are often investigated under the assumption of an infinite slab and real effects arising due to the confinement and boundaries are ignored. A nanoflake is a monolayer with spatial dimensions less than 100 nm. The structural, electronic and optical properties of such nanoflakes may be strongly influenced by varying their lateral size.

We study the MoS$_{2}$ monolayer nanoflakes for two commonly known edge structures, zigzag and armchair, to investigate the stable edge structure for smaller nanoflakes. Structures before geometry relaxation without any edge termination are shown in Fig.~\ref{fig:Edges}. Zigzag structures have double-coordinated, bridge-like S or Mo atoms on the edges [Fig.~\ref{fig:Edges}(a)], whilst armchair have single-coordinated, antenna-like S or Mo atoms [Fig.~\ref{fig:Edges}(b)]. We relaxed both of these types of structure, encountering convergence issues for the two larger structures (72 atoms and 105 atoms). We succeeded in getting convergence of $<10^{-7}$ RMS change in the density matrix for the 72-atom structures in both zigzag and armchair edge configurations. For the 105-atom structure, we obtained convergence of $<10^{-5}$ RMS change in the density matrix in zigzag edge configuration, but could not converge the 105-atom armchair edge configuration at all. This therefore, sets the maximum structure size in our calculations. In \textsc{gaussian09}, the energy change is not a criterion for convergence, however, the worst level of convergence for the largest structure, \textit{i.e.}, $<10^{-5}$ RMS change in density matrix, typically corresponds to $<10^{-10}$ Ha change in energy \cite{gaussiao09}. For the larger structures, we are more confident of the trends instead of the absolute values of energy.

The ground-state energies as functions of the size of the nanoflakes are shown in Fig. \ref{fig:energetics}. Assuming that the edge width remains constant for any flake size, as the flakes get larger the ratio of number of edge atoms to core atoms decreases significantly because the number of core atoms increases more rapidly. (A quick circular approximation shows the core area $\propto L^{2}$, whilst treating the edge as an annulus gives area $\propto L$, where $L$ is the radius of the core.)  The structure becomes more stable as it becomes larger. Fig. \ref{fig:energetics} shows that zigzag is the most stable configuration for nanoflakes of size less than 2 nm. Out of the trial fit functions (we expect a function that decreases with the increase in the flake size and finally asymptotes to a limiting value to physically describe our model) $A_{k}{\rm exp}(-B_{k}N)+C_{k}$, $\frac{D_{k}}{N} +F_{k}$, and $\frac{G_{k}}{N^{2}}+H_{k}$, the exponential function fits our data best. The subscript $k$ indexes various properties as discussed for various figures. Here $k = 1$ for zigzag structures and $k=2$ for armchair. $N$ is the number of atoms. The parameters $A_{k}$, $B_{k}$, and $C_{k}$ are solved through the least-squares curve-fitting method. For zigzag structures, we found these parameters; $A_{1} = 2.9$, $B_{1} = 0.0355$ and $C_{1} = -2.384\times 10^{3}$ and for armchair; $A_{2} = 3.75$, $B_{2} = 0.06$ and $C_{2} = -2.38\times 10^{3}$. We extrapolate the fit function to generalize the behaviour for larger nanoflakes of size up to 200 atoms. We find that the zigzag-edged structure is always more stable than the armchair configuration. All further properties are discussed for zigzag edge configuration only because it is the stablest.

The relaxed structures of MoS$_{2}$ monolayer nanoflakes are shown in Fig.~\ref{fig:core_shell}. We compared the atomic positions in the relaxed structures with their unrelaxed positions in the bulk structure \cite{Dickinson1923}. The colour of the atoms in this figure is proportional to the displacement of atoms from their bulk positions, $\Delta R$ as defined in Eq.~\ref{Eq:Colour}, with $i$ indexing all the atoms in the flakes.

The smaller nanoflakes are strongly distorted after relaxation compared to their unrelaxed structures except for the 9-atom flake. In the smallest structure having 9 atoms, all the Mo atoms are unsaturated symmetrically and all of them show the same distortion with a mean Mo--Mo length of 2.52 \r{AA}, while in the bulk structure this length is reported to be 3.15 \r{AA} \cite{Dickinson1923}. Similarly all the S atoms show the identical distortion with S--S lengths of 3.43 \r{AA}. For the 24-atom structure, maximum distortion is observed at the acute-Mo [a(Mo)] corner. This maximum Mo--Mo length is shown by red-arrowed line $d1$ in Fig.~\ref{fig:core_shell}(b), and is 2.66 \r{AA}. As we move to the next structure (45 atoms), this maximum distortion is shifted to the two obtuse-Mo \& S [o(Mo \& S)] corners. The unsaturated Mo atoms showing maximum distortion are displaced inwards [Fig.~\ref{fig:core_shell}(c)]; for example, $d2$ and $d3$ are shortened to 2.50 \r{AA} while in the bulk structure, they are 3.15 \r{AA}. The maximum S--S length distortion in the same structure is $d4$ = 3.29 \r{AA}. As the structures get larger, we observe that the central zones show greatly reduced variation [Fig.~\ref{fig:core_shell}(f)] after the optimization. For the  two larger structures (with 72 and 105 atoms), the maximum distortion is shifted towards the acute-S [a(S)] corner ring [Fig.~\ref{fig:core_shell}(d-e)]. Both of these structures show identical geometric behaviour and the maximum distortions are on the Mo--Mo lengths shown by red-arrowed lines $d5$ = $d6$ = 2.60 \r{AA}. These two structures show a well-established core whose mean structural parameters approach the bulk structure values \cite{Dickinson1923}. 

We have done an analysis of the Mo--S bond lengths in the central zones of our relaxed structures and compared them with the bulk Mo--S bond lengths of 2.41 $\pm$ 0.06 \r{AA} reported in \cite{Dickinson1923}. Fig. \ref{fig:core_shell}(f) shows the percentage variation of the mean Mo--S bond lengths in the central zone of each structure with the bulk Mo--S bond length, $\Delta r_{\rm Mo-S}$ defined as:
\begin{equation}
\Delta r_{\rm Mo-S} = \frac{r_{\rm Mo-S}^{\rm flake} -r_{\rm Mo-S}^{\rm bulk} }{r_{\rm Mo-S}^{\rm bulk}} \times 100 \%
\end{equation}

The error bars show the range of the minimum and maximum bond lengths in the central zone from the mean value. The smallest flake shows minimum mismatch from the bulk bond lengths. The flake with 24 atoms shows a mean mismatch of 5\% from the bulk values. After that as the flake size increases, this percentage mismatch from the bulk values declines and then converging to a value of 2\% [Fig.~\ref{fig:core_shell}(f)] for the two larger structures. 

\section{Size-dependent electronic properties}

\begin{figure}[tb!]
\centering
\includegraphics[scale=0.8]{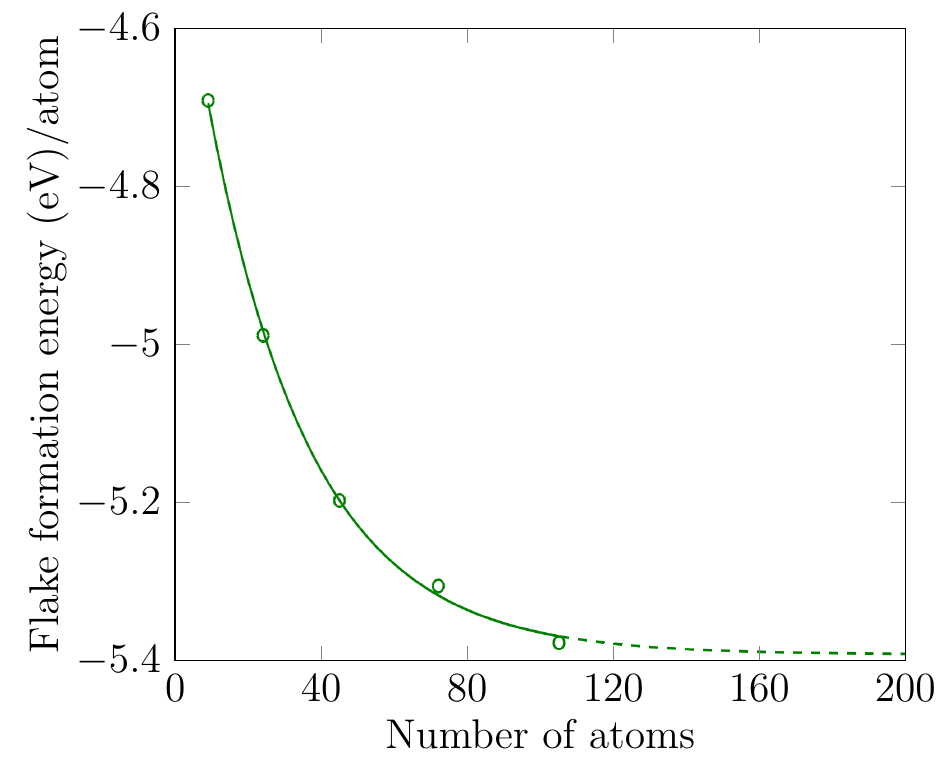}
\caption{Flake formation energy as a function of nanoflake size. As the size increases, the formation energy decreases. The solid line is an exponential fit to the data. An extrapolation for larger structures is shown by the dashed line.}
\label{fig:FFE}
\end{figure}

To indicate the stability and the tendency of flakes to grow, we calculated the size-dependent flake-formation energy (FFE) of MoS$_{2}$ monolayer nanoflakes given by 
\begin{equation}
{\rm FFE} = E_{\rm flake}(\text {Mo}_{n}\text S_{2n})-nE(\text {Mo}) -2nE(\text S),
\end{equation}
where \textit{n} is the number of Mo atoms and 2\textit{n} the number of S atoms in the flake, \textit{E}(Mo) is the energy of a single Mo atom, \textit{E}(S) is the energy of a single S atom, and $ E_{\rm flake}(\text {Mo}_{n}\text S_{2n})$ is the energy of the flake having \textit{n} Mo atoms and 2\textit{n} S atoms. As defined, FFE $<$ 0 indicates that the flake is more stable than its constituent atoms. Figure \ref{fig:FFE} shows that with the increase in nanoflake size the FFE decreases sharply, so more energy is released by adding atoms in the larger flakes indicating that the flakes tend to grow energetically. Conversely, more energy is required to break the larger flakes into their constituents.

We again fit an exponential function $A_{k}{\rm exp}(-B_{k}N)+C_{k}$ for similar reasons as those applied regarding Fig.~\ref{fig:energetics} to fit the data and extrapolated this fit to generalize the behaviour for the larger flakes as shown in Fig.~\ref{fig:FFE}. Here $k =3$ for FFE and $A_{3} = 0.9394$, $B_{3} = 0.0355$, and $C_{3} = -5.3914$ are the parameters solved through least-squares curve fitting. 

\begin{figure}[tb]
\centering
\includegraphics[scale=0.8]{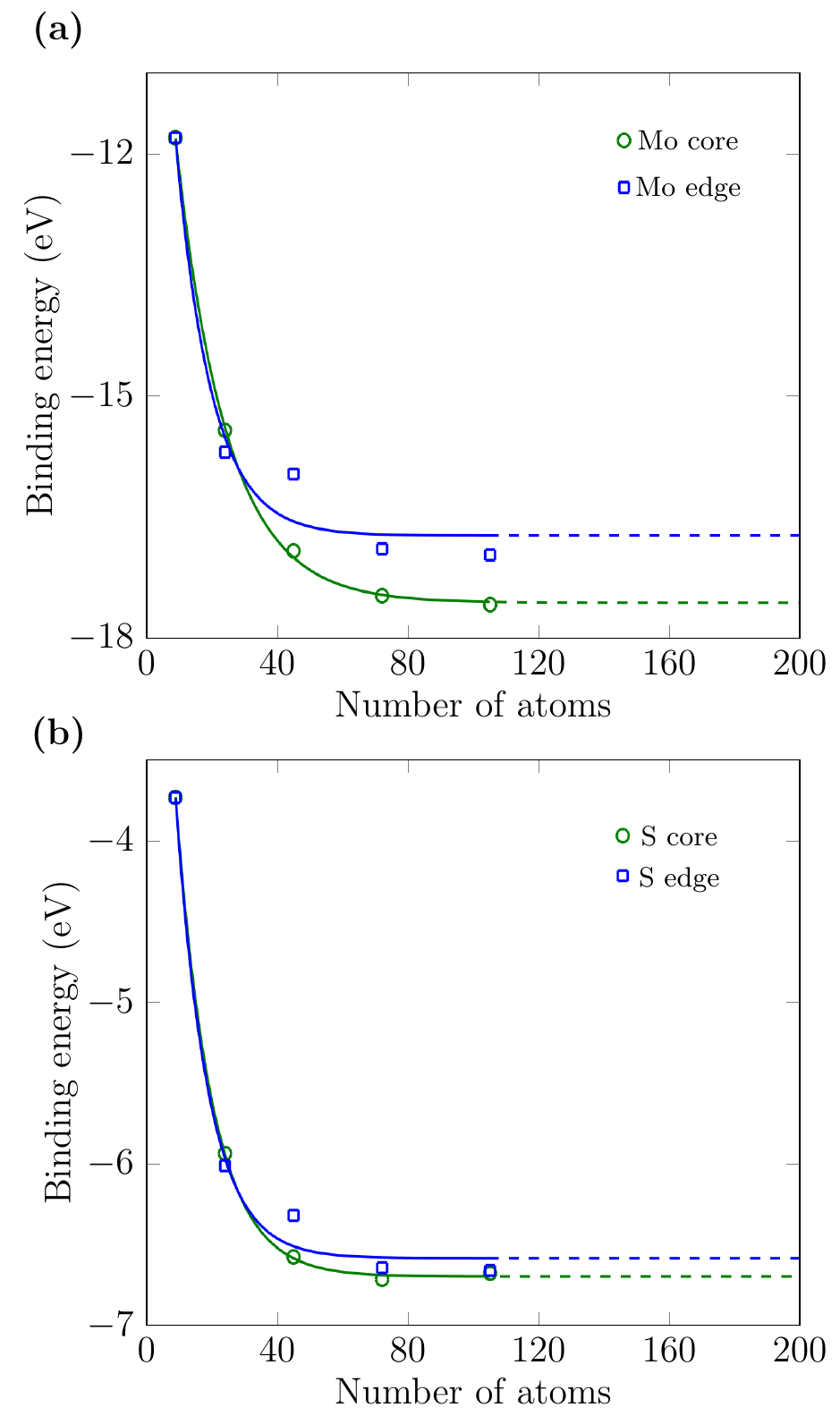}
\caption{(a) Binding energies of Mo atoms as functions of number of atoms in the flakes. (b) Binding energies of S atoms as functions of size of the nanoflakes. Solid lines in both (a) and (b) are exponential fits to the data and extrapolated to predict the behaviour for the larger flakes shown by the dashed lines.}
\label{fig:BE}
\end{figure}

We calculated the binding energies for all flake sizes and present them as a function of size in Fig.~\ref{fig:BE}. We removed a Mo/S atom from as close as possible to the centre of the core or the edge as possible. The binding energy for the Mo atoms is given by
\begin{equation}
E_{B_{\rm Mo}} = E(\text {Mo}_{n}{\text S}_{2n}) - E(\text {Mo}_{n-1}\text S_{2n}) - E(\text{Mo}).
\end{equation}
Similarly, the binding energy for S atoms is given by
\begin{equation}
 E_{B_{\text S}} = E({\rm Mo}_{n}\text S_{2n}) - E(\text {Mo}_{n}\text S_{2n-1}) - E(\text S).
\end{equation}

Negative values of the binding energy indicate that energy is required to remove an atom from a nanoflake. The negative dependence with size means that the cost rises with flake size. For example, removing a Mo atom from the core of a 45-atom flake requires $\thicksim$1.2 eV more energy than removing it from the core of a 24-atom flake. $E_{B}=-E_{D_{\rm form}}$, where $E_{D_{\rm form}}$ is the defect-formation energy so we can also calculate the energy required to create a Mo or S vacancy in the core or on the edge of the nanoflakes. From Fig.~\ref{fig:BE}, significantly more energy is required to create a Mo vacancy as compared to a S vacancy. Also there is no major difference in the energy required to create a Mo vacancy in the core or in the edge in smaller flakes but as the size of the flakes increases, comparatively it becomes easier for defects to form on the edges. In case of S atoms, approximately the same energy is required to create a S vacancy in the core or in the edge as shown in Fig. \ref{fig:BE}(b). 

We again used the exponential function $A_{k}{\rm exp}(-B_{k}N)+C_{k}$ to fit the data. Here $k = 4$ for Mo core, $k = 5$ for Mo edge, $k = 6$ for S core, and $k = 7$ for S edge binding energies. The parameters solved through least-squares-curve fitting are as: for Mo core binding energies, $A_{4} = 8.80$, $B_{4} = 0.06$, and $C_{4} =-16.61 $; for Mo edge binding energies, $A_{5} = 9.62$, $B_{5} = 0.09$, and $C_{5} =-15.90 $; for S core binding energies, $A_{6} = 6.69$, $B_{6} = 0.09$, and $C_{6} =-6.6 $; and for S edge binding energies, $A_{7} = 7.07$, $B_{7} = 0.10$, and $C_{7} =-6.5 $. We again generalize these defect formation energies for flakes larger than 105 atoms using an exponential fit. Fits for Mo edge and S edge binding energies are insufficient to predict binding energies for larger structures due to the 45-atom structure values.

\begin{figure*}[!]
\centering
\includegraphics[width=18cm, scale=1]{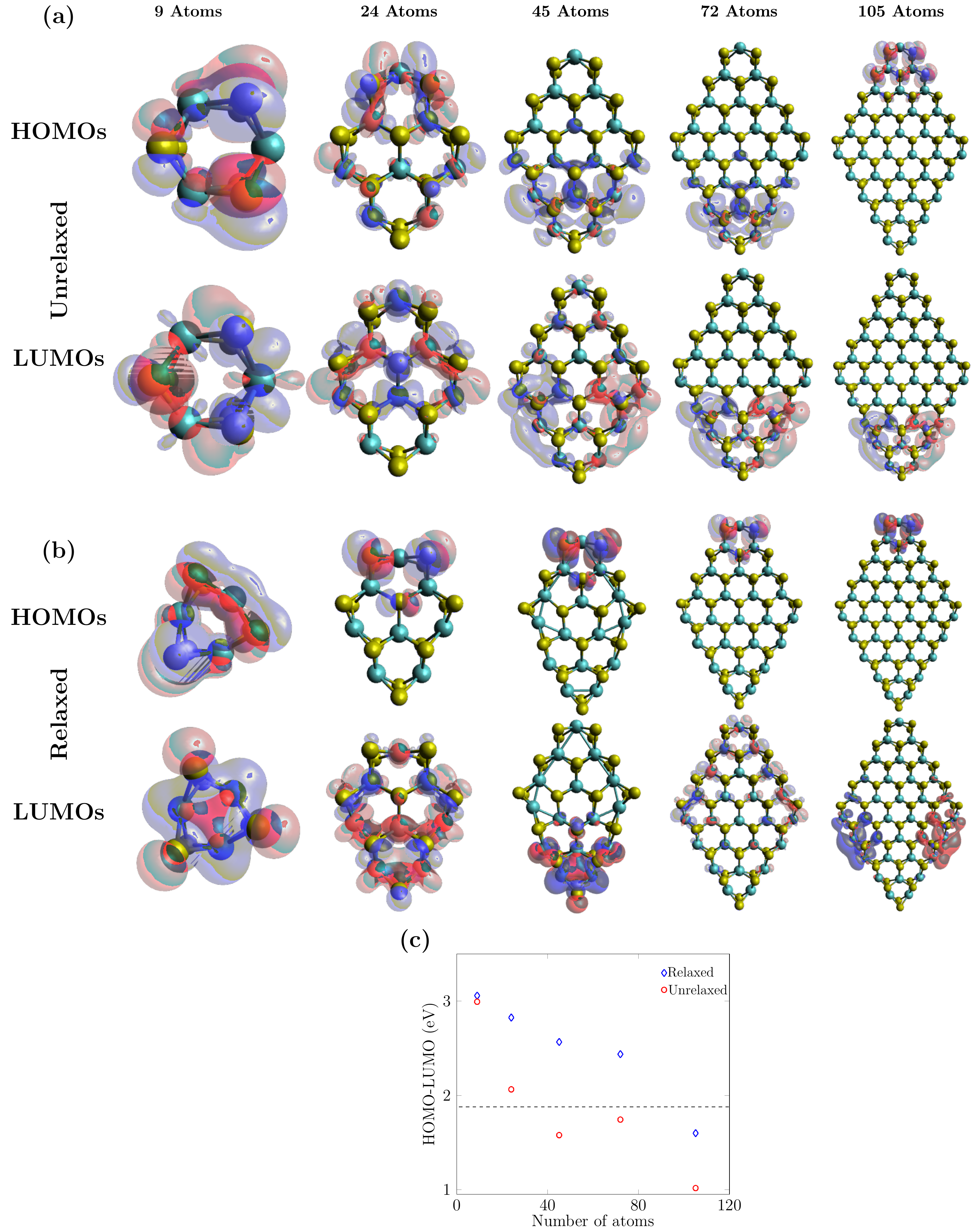}
\caption{HOMO and LUMO charge densities of (a) unrelaxed, and (b) relaxed zigzag nanoflakes for various flake sizes at an isosurface value of 0.02 $e/\text{Bohr}^{3}$. (c) HOMO-LUMO gaps as functions of size of the nanoflakes for both unrelaxed and  relaxed structures. As the size of the nanoflakes increases, the gaps generally decrease. The black-dashed line indicates the known experimental band gap for an infinite sheet of MoS$_{2}$ monolayer \cite{PhysRevLett}.}
\label{fig:HOMO_LUMO}
\end{figure*}

\begin{table*}[!]
\newcolumntype{P}[1]{>{\centering\arraybackslash}p{#1}}
\newcolumntype{M}[1]{>{\centering\arraybackslash}m{#1}}
\caption{A comparison of the mean lengths in the relaxed, passivated structures with and without H dimers on the Mo corner rings, encircled by green on all the structures in Fig.~\ref{Fig:Termination}. There is a maximum mismatch of 2\% in the S--S length in the 9 atom structure.}
\centering
\begin{tabular}{ M{1.5cm} M{1.5cm} M{1.5cm} M{1.5cm} M{1.5cm} M {1.5cm} M{1.5cm}}\\
\toprule
{}& \multicolumn{2}{c}{Mean Mo--Mo (\r{AA})} &\multicolumn{2}{c}{Mean S--S (\r{AA})}&\multicolumn{2}{c}{Mean Mo--S (\r{AA})}\\
\midrule \hline
Nanoflake size & with H dimer & without H dimer& with H dimer & without H dimer& with H dimer & without H dimer\\
\midrule \hline
9 atoms&  2.41&2.40 & 3.65&3.74&2.60&2.60  \\
24 atoms& 2.73&2.69&3.48&3.52&2.53&2.54 \\
45 atoms&2.72&2.69&3.51&3.55&2.53&2.53 \\
72 atoms&  2.73&2.70&3.52&3.54&2.53&2.53\\
 \bottomrule
\end{tabular}
\label{table:Bondlengths}
\end{table*}

To predict the electronic properties of ultra-small MoS$_{2}$ monolayer nanoflakes, we calculated their HOMO-LUMO gaps and charge densities of their HOMO and the LUMO (Fig.~\ref{fig:HOMO_LUMO}). With an increase in flake size, the HOMO-LUMO gap decreases for both unrelaxed and relaxed structures which is in keeping with intuition around the increase in the HOMO-LUMO gap with decreasing particle size as discussed in the methdology section. The experimentally obtained band gap for the infinite MoS$_{2}$ monolayer structure is 1.88 eV \cite{PhysRevLett} shown by the dashed line in Fig.~\ref{fig:HOMO_LUMO}. For larger flakes, we have not observed the band gap converging to this value. One possible cause could be dangling bonds in the nanoflakes. To address this, we study passivated structures in the next section.

To get deeper insight into the HOMO-LUMO behaviour as a function of nanoflake size, we calculated charge-density plots (Fig.~\ref{fig:HOMO_LUMO}) for structures before and after the geometry relaxation. We can see that the majority of the HOMO and the LUMO charge densities are lying on the corners and edges in all of these structures except the 9-atom nanoflake where they are scattered over the whole structure. No single, stand-out trend is observed in all the structures. In short, the charge density is highly sensitive to the structural size for these small sized nanoflakes.

\section{Hydrogen Passivation of molybdenum-disulphide nanoflakes}
\begin{figure*}[tb]
\centering
\includegraphics[scale=0.5]{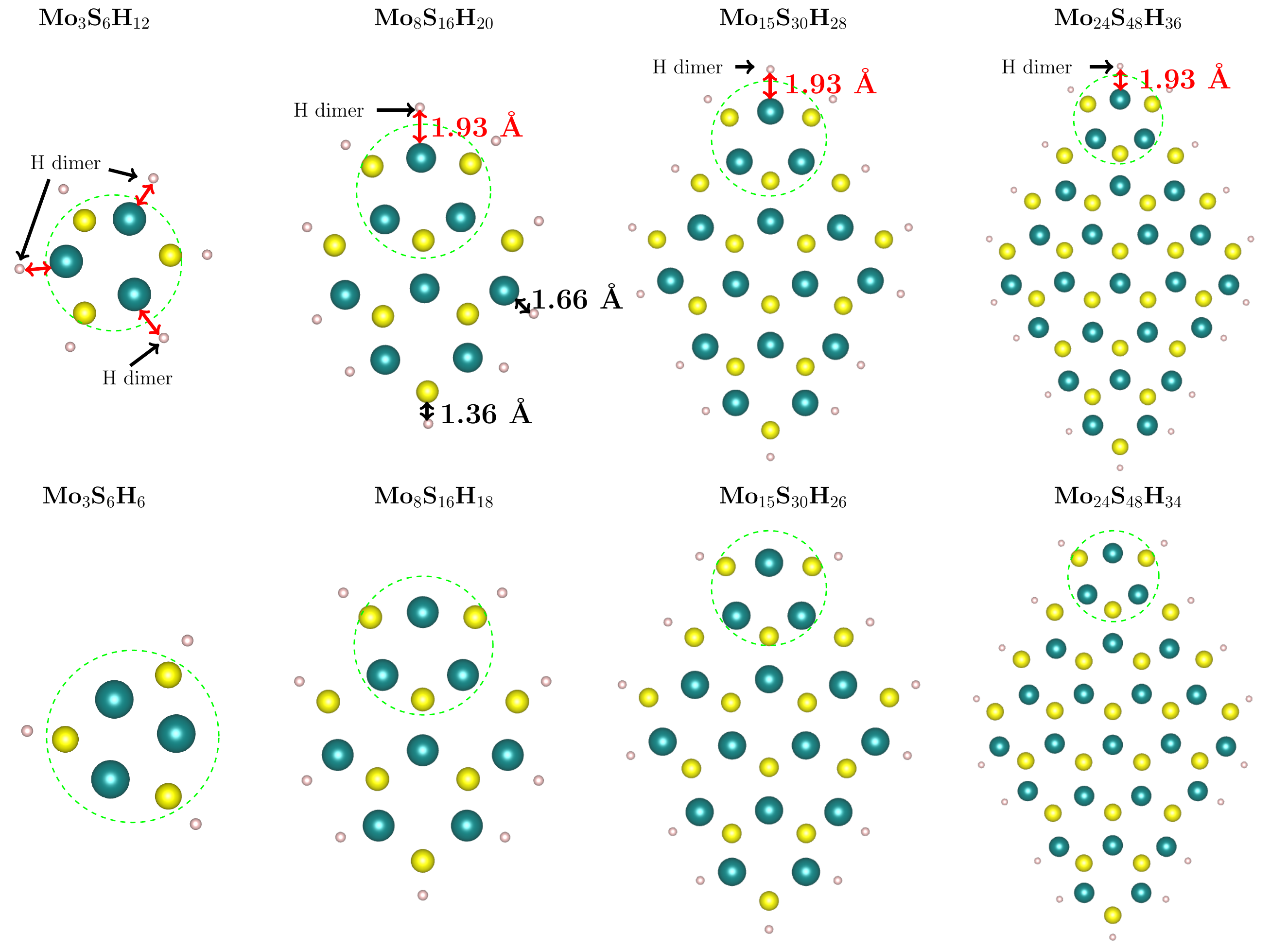}
\caption{MoS$_{2}$ monolayer nanoflakes passivated with H atoms and relaxed. Each Mo edge atom is passivated with 2 H atoms and each S edge atom is passivated with one H atom. The top row shows nanoflakes with H dimers not bonded to the flake (labelled). We removed these H dimers, relaxed the nanoflakes again and the relaxed structures are shown in the bottom row. The mean Mo--Mo, S--S lengths and Mo--S bond lengths in the green-encircled ring of each flake are reported in Table \ref{table:Bondlengths}.}
\label{Fig:Termination}
\end{figure*}

\begin{figure}[b!]
\centering
\includegraphics[scale=0.85]{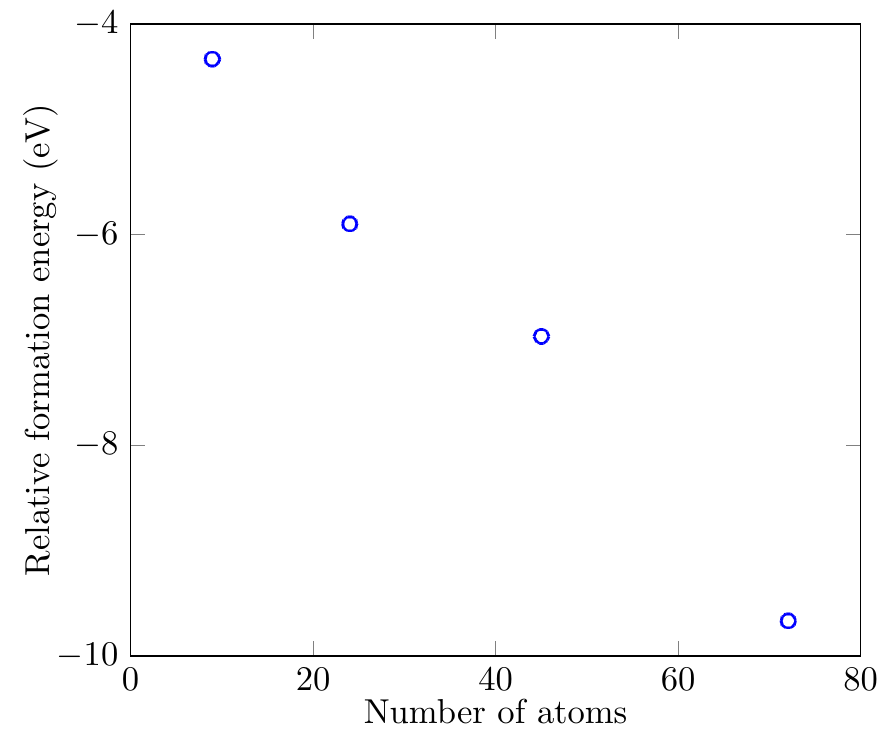}
\caption{Energy difference between the passivated and unpassivated structures. The passivated structures are significantly more stable than the unpassivated ones in all cases.}
\label{Fig:Stability}
\end{figure}

Dangling bonds exist on the edges and corners of the nanoflakes. The smallest structure with 9 atoms has no fully coordinated atoms. The structure with 24 atoms possesses 5 under-coordinated Mo and 10 under-coordinated S atoms. Similarly, the structures with 45, 72, and 105 atoms possess 7 Mo and 14 S, 9 Mo and 18 S, and 11 Mo and 22 S under-coordinated atoms respectively. 

It has been reported that the edge Mo atoms with unsaturated bonds may not be stable \cite{Halveg2000,Lauritsen2007}. Also in \cite{TOPSOE1993}, Topsoe \textit{et al.} have reported the presence of S--H groups on the edges of MoS$_{2}$ clusters experimentally. In \cite{Passivation}, Loh \textit{et al.} have also passivated the S with H atoms in their triangular MoS$_{2}$ quantum dot on hexagonal boron nitride substrate. 

To understand the effects of dangling bonds on the properties of the structures, we passivated both Mo and S edges with H atoms. We passivated each edge Mo atom with 2 H atoms as we expect Mo atoms to be bonded with 6 atoms in this particular MoS$_{2}$ stoichiometry. We also tested single H-termination of all edge Mo atoms and could not obtain converged, relaxed structures. We suspect this means that such structures are energetically unfavourable. We terminated each edge S atom with one H atom as all the central S atoms form three bonds with their neighbouring Mo atoms. We relaxed these passivated structures and observed that on the acute-Mo corner of all the nanoflakes, the H atoms are pushed away and they do not appear to bond to Mo atoms (Fig. \ref{Fig:Termination}). We investigated this non-bonding of corner Mo atoms with H atoms by checking their bond lengths. The average Mo--H bond length for all the edge Mo atoms is 1.665 $\pm$ 0.005 \r{AA} while on the corner it is 1.94 \AA. The two H atoms on the Mo corner have H--H bond length of 78 pm. We calculated the H--H bond length in a lone H dimer as 74 pm which is in good agreement with the known value \cite{Gorton}. The H--H bond length value, \textit{i.e.}, 78 pm on the acute-Mo corner in all passivated flakes is close enough to the known H--H value that we can believe that they are making a separate H$_{2}$ molecule. 

We removed the corner H atom, relaxed the structures again and observed almost the same structural parameters on the corner as with the corner H atoms. We compared the mean Mo--Mo, S--S, and Mo--S lengths of the acute-Mo corner ring (encircled by green in Fig.~\ref{Fig:Termination}) in Table \ref{table:Bondlengths} for the relaxed structures with and without the H dimer on the corner Mo atom. For all the structures, there is a minimal change in the bond lengths between 0--2\%. All the S atoms bond well to one H atom each with an average S--H bond length of 1.365 $\pm$ 0.005 \r{AA}. We could not obtain a relaxed, converged 105-atom (we are not counting the number of H atoms to keep the number of atoms in each flake consistent with the previous discussion) passivated structure.

To calculate the stability, we have compared the energies of the passivated structures with the corresponding unpassivated ones. We found that the passivated structures are significantly more stable than the unpassivated ones by 4.33, 5.9, 6.96, and 9.66 eV for 9, 24, 45, and 72 atoms respectively as shown in Fig.~\ref{Fig:Stability} where the relative formation energy (RFE) is;  

\begin{equation}
\text{RFE} = E(\text{Mo}_{n}\text{S}_{2n}\text{H}_{m}) - E(\text{Mo}_{n}\text{S}_{2n}) - \frac{m}{2}\text{H}_{2}, 
\end{equation}
where $m$ is the number of H atoms in the passivated structures.

\begin{figure}[htb]
\centering
\includegraphics[scale=0.8]{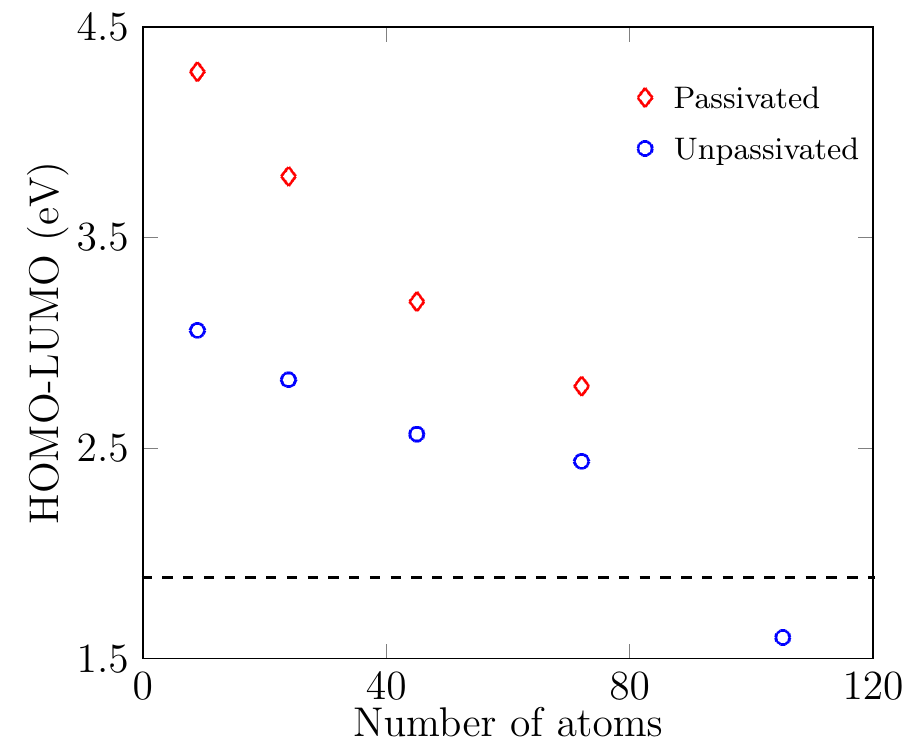}
\caption{HOMO-LUMO gap of the unpassivated structures (blue circles) versus the passivated structures (red diamonds). Passivated structures have larger HOMO-LUMO gaps. The black-dashed line indicates the known experimental band gap of infinite MoS$_{2}$ monolayers as reported in \cite{PhysRevLett}.}
\label{Fig:HL_Ter}
\end{figure}

Passivation of the dangling bonds modifies the electronic structure, charge densities and hence the HOMO-LUMO gap. In Fig.~\ref{Fig:HL_Ter}, the HOMO-LUMO gap of the passivated structures is contrasted against the unpassivated ones. We find that the HOMO-LUMO gap widens with passivation. We suspect this is because of the removal of dangling bonds. This effect is significant in smaller nanoflakes but as the size increases, the ratio of edge to core atoms decreases. Hence, due to fewer edge states in the larger structures, the HOMO-LUMO gap difference (both relative and absolute) between the passivated and the unpassivated structures becomes smaller. 

\begin{figure*}[!]
\centering
\includegraphics[scale=0.65]{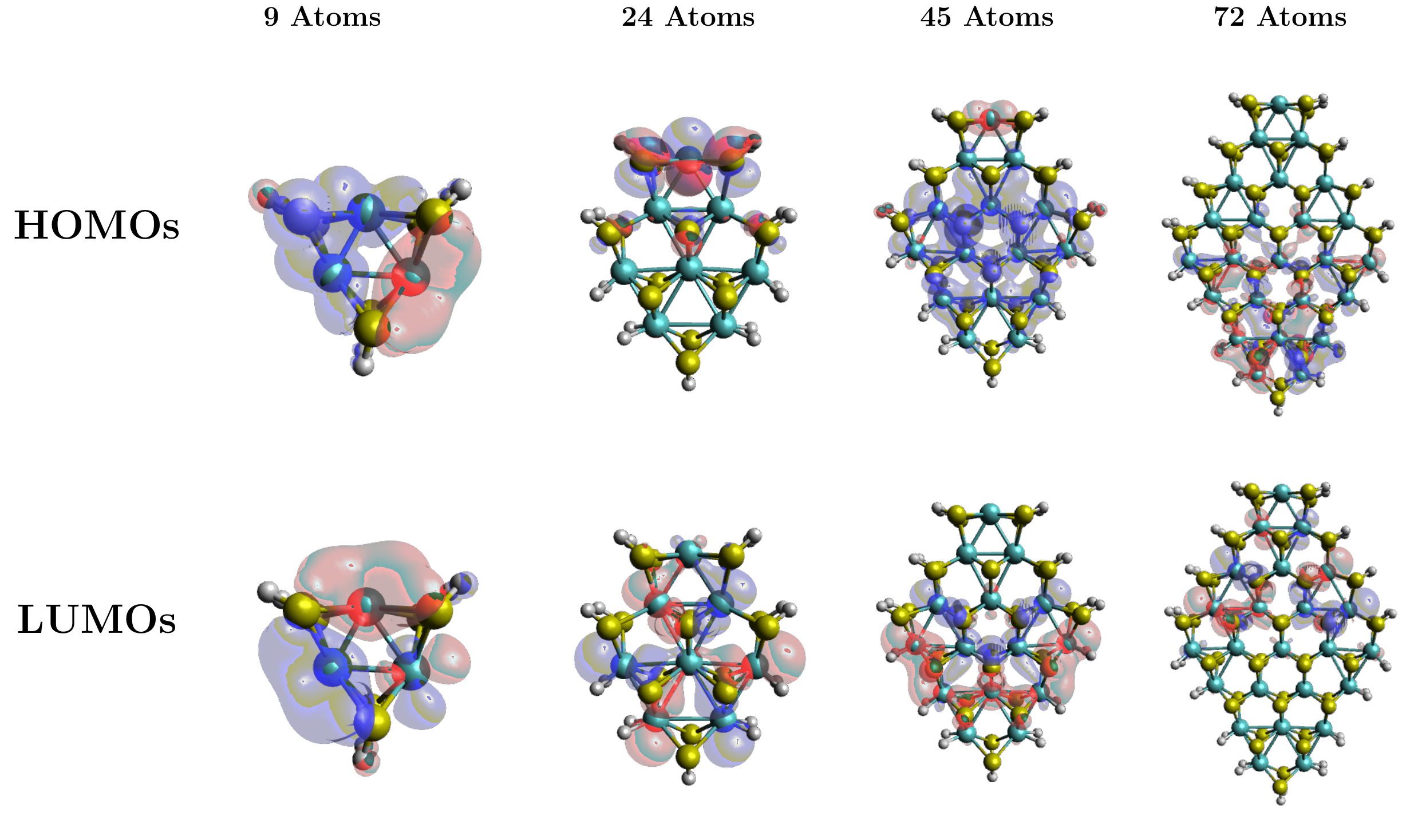}
\caption{Charge densities in the HOMOs and LUMOs of the passivated structures for various sizes of nanoflakes for an isosurface value of 0.02 $e/\text{Bohr}^{3}$. Charge densities are very sensitive to the size of the nanoflakes.}
\label{Fig:CHG_Ter}
\end{figure*}

The charge densities of the passivated structures are shown in Fig.~\ref{Fig:CHG_Ter}. These are much more distributed states in contrast to the charge density plots for unpassivated, relaxed structures [Fig.~\ref{fig:HOMO_LUMO}(b)]. Thus passivation makes HOMO/LUMO states in these small-sized flakes more like the expected infinite monolayer. 

\section{Conclusions}

In summary, we have investigated the size-dependent structural and electronic properties of MoS$_{2}$ monolayer nanoflakes of sizes up to 2 nm using DFT. Our main focus has been to explore the small-sized nanoflakes. We provide more-detailed information for engineering small-sized nanoflakes by reporting the energetically favourable edge configuration and size of the nanoflakes. We predicted the trends in the energetics as functions of size. We passivated the structures to explore the effects of passivation on small-sized nanoflakes. We found the passivated structures to be more stable, with wider HOMO-LUMO gaps than unpassivated ones. We observe several strong size dependencies of various properties.

The size-dependence of the HOMO-LUMO gap of these small-sized nanoflakes holds promise for opto-electronic applications. However, due to the size-dependent energetics involved, one must take care in the manufacture/selection of these flakes. Due to limited computational resources, we were able to model only small-sized nanoflakes and can predict trends for larger flakes only by extending the fit functions. However, an extension of the current work to nanoflakes larger than 2 nm would be a good benchmark for the DFTB size-dependent HOMO-LUMO gaps reported by Wendumu \textit{et al.} \cite{Wendumu2014}.
\section{Acknowledgements}
The authors acknowledge financial support from the Australian Research Council (Project Nos. DP130104381, CE140100003, FT160100357, LE160100051, CE170100026). This work was supported by computational resources provided by the Australian Government through the National Computational Infrastructure (NCI) under the National Computational Merit Allocation Scheme. The authors thank Rika Kobayashi (NCI) for useful discussion and advice. 

\appendix
\section{}

Here we present a brief overview of the analysis of different functionals on a small MoS$_{2}$ monolayer nanoflake having 9 atoms. We also show energy-level diagrams for the passivated and unpassivated structures of various sizes to study the size-dependence of MoS$_{2}$ monolayer nanoflakes and the effects of passivation on the electronic energy levels of the nanoflakes.

To choose the appropriate functional for modelling these small-sized nanoflakes, we made a comparison of the HOMO-LUMO gap using different functionals in \textsc{gaussian09} as shown in Table \ref{table:Tab3}. We faced an energy convergence issue when using the BP86 \cite{Becke88,BP86} functional and did not use it for further modelling as we suspected that the convergence issues would be worse for larger flakes using this functional. For HSEH1PBE \cite{HSEH1PBE_1,HSEH1PBE_2,HSEH1PBE_3,HSEH1PBE_4,HSEH1PBE_5,HSEH1PBE_6}, B3LYP \cite{Becke88, ECLYP, B3LYP}, PBE1PBE \cite{PBE1PBE}, B3PW91 \cite{Becke88,B3PW91}, PBEh1PBE \cite{PBEh1PBE}, and M05 \cite{M05}, we obtained gaps smaller than the known experimental band gap in infinitely large sheet of MoS$_{2}$ monolayer. We expect the HOMO-LUMO gap to decrease with increasing flake size and then converge to the infinite monolayer MoS$_{2}$ band gap for larger flakes as discussed in the main paper. Thus for these functionals, we expect the results to get worse with any increase in flake size. The M052X \cite{M052XFunctional} and BHandHLYP \cite{Functional} functionals predicted reasonable gaps for this small nanoflake and we can conjecture that they might asymptote near the experimental value for larger flakes.

\begin{figure*}[!]
\centering
\includegraphics[scale=1]{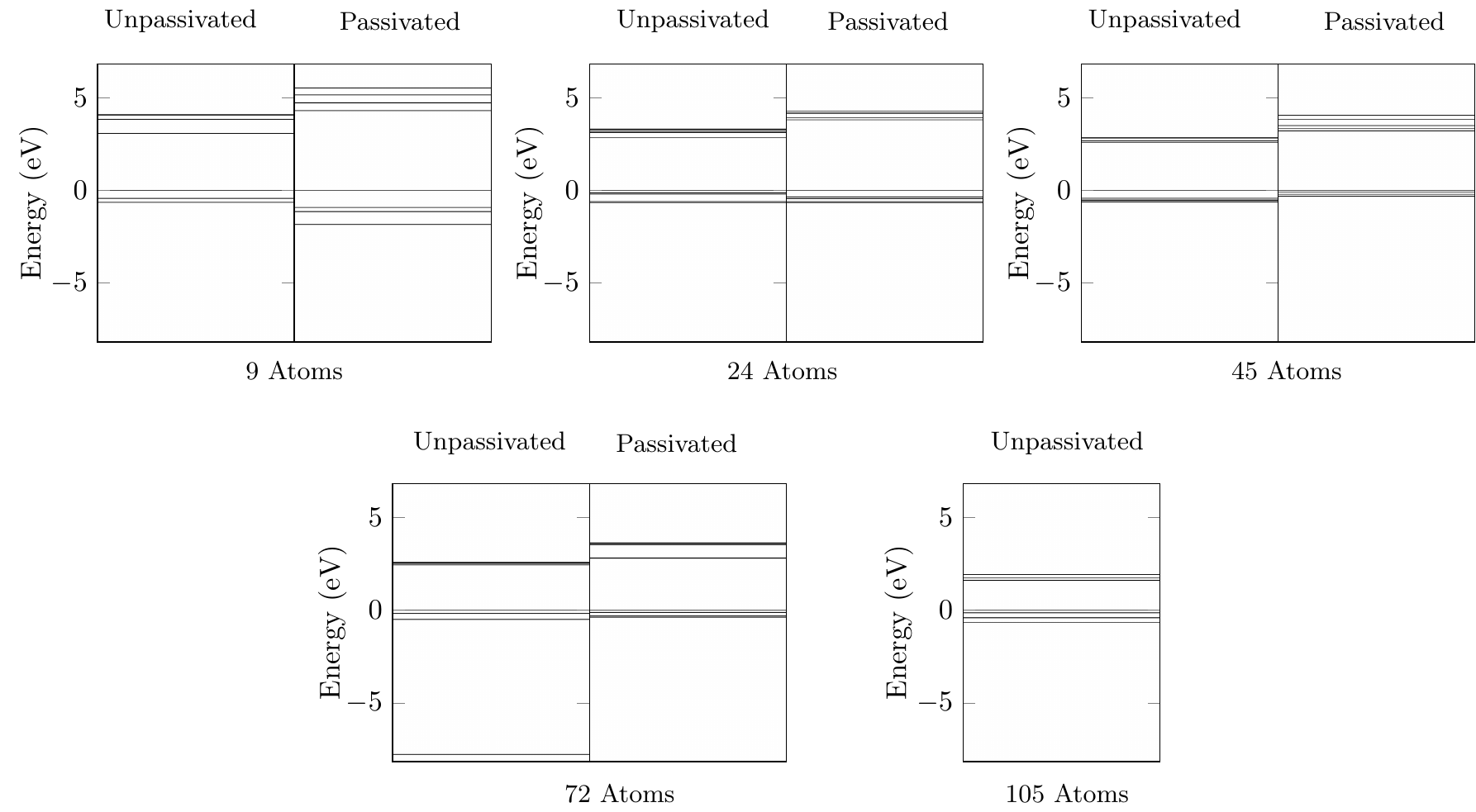}
\caption{Energy levels from HOMO-4 to LUMO+4 in the unpassivated and passivated nanoflakes of various sizes. The HOMOs are scaled to zero on the energy axes.}
\label{Fig: EnergyLevels}
\end{figure*}

In Fig.~\ref{Fig: EnergyLevels}, we have shown the several energy levels from HOMO-4 to LUMO+4 for both passivated and unpassivated nanoflakes. The HOMO is scaled to zero on the energy axes for all flakes. In both passivated and unpassivated flakes, the HOMO-LUMO gap shrinks with increasing size as discussed in the main paper. In the unpassivated structures, from 9 atoms to 72 atoms, the conduction band gets significantly denser with increasing size, while there is no significant change in the valence band's level spacing. For 105 atoms, the level spacing in the conduction band increases slightly again. 

In the passivated structures, the valence bands get denser with increasing flake size while oscillating behaviour is observed in the conduction bands, which first gets denser from 9 atoms to 24 atoms, then slightly splits again for 45 atoms and then becomes denser again for 72 atoms. For all these structures, the HOMO-LUMO gap gets wider after passivation which is consistent with the idea that dangling bonds widen the band gap discussed in the main paper.

In summary, we obtained a reasonable subset of functionals to use for further modelling. We also found that the energy levels are very sensitive to the nanoflake sizes and that the dangling bonds play an important role in the HOMO-LUMO gap.

\begin{table}[h!]
\newcolumntype{P}[1]{>{\centering\arraybackslash}p{#1}}
\newcolumntype{M}[1]{>{\centering\arraybackslash}m{#1}}
\centering
\caption{An analysis of the HOMO-LUMO gap in \textsc{gaussian09} for a 9-atom nanoflake under different functionals. }
\begin{tabular}{ M{3.5cm} M{3.5cm} }
 \hline
 Functionals     & HOMO-LUMO gap (eV)\\
 \hline
 B3LYP&  0.75  \\
 BHandHLYP & 3.06 \\
 HSEH1PBE  & 0.25  \\
BP86&   Convergence error \\
B3PW91 & 1.44 \\
PBE1PBE & 1.73\\
PBEh1PBE & 1.70\\
M05 & 0.67 \\
M052X & 3.27\\
 \hline
\end{tabular}
\label{table:Tab3}
\end{table}

\bibliography{MoS2Ref}

\begin{thebibliography}{49}%
\makeatletter
\providecommand \@ifxundefined [1]{%
 \@ifx{#1\undefined}
}%
\providecommand \@ifnum [1]{%
 \ifnum #1\expandafter \@firstoftwo
 \else \expandafter \@secondoftwo
 \fi
}%
\providecommand \@ifx [1]{%
 \ifx #1\expandafter \@firstoftwo
 \else \expandafter \@secondoftwo
 \fi
}%
\providecommand \natexlab [1]{#1}%
\providecommand \enquote  [1]{``#1''}%
\providecommand \bibnamefont  [1]{#1}%
\providecommand \bibfnamefont [1]{#1}%
\providecommand \citenamefont [1]{#1}%
\providecommand \href@noop [0]{\@secondoftwo}%
\providecommand \href [0]{\begingroup \@sanitize@url \@href}%
\providecommand \@href[1]{\@@startlink{#1}\@@href}%
\providecommand \@@href[1]{\endgroup#1\@@endlink}%
\providecommand \@sanitize@url [0]{\catcode `\\12\catcode `\$12\catcode
  `\&12\catcode `\#12\catcode `\^12\catcode `\_12\catcode `\%12\relax}%
\providecommand \@@startlink[1]{}%
\providecommand \@@endlink[0]{}%
\providecommand \url  [0]{\begingroup\@sanitize@url \@url }%
\providecommand \@url [1]{\endgroup\@href {#1}{\urlprefix }}%
\providecommand \urlprefix  [0]{URL }%
\providecommand \Eprint [0]{\href }%
\providecommand \doibase [0]{http://dx.doi.org/}%
\providecommand \selectlanguage [0]{\@gobble}%
\providecommand \bibinfo  [0]{\@secondoftwo}%
\providecommand \bibfield  [0]{\@secondoftwo}%
\providecommand \translation [1]{[#1]}%
\providecommand \BibitemOpen [0]{}%
\providecommand \bibitemStop [0]{}%
\providecommand \bibitemNoStop [0]{.\EOS\space}%
\providecommand \EOS [0]{\spacefactor3000\relax}%
\providecommand \BibitemShut  [1]{\csname bibitem#1\endcsname}%
\let\auto@bib@innerbib\@empty
\bibitem [{\citenamefont {Novoselov}\ \emph {et~al.}(2005)\citenamefont
  {Novoselov}, \citenamefont {Geim}, \citenamefont {Morozov}, \citenamefont
  {Jiang}, \citenamefont {Katsnelson}, \citenamefont {Grigorieva},
  \citenamefont {Dubonos},\ and\ \citenamefont {Firsov}}]{Novoselove}%
  \BibitemOpen
  \bibfield  {author} {\bibinfo {author} {\bibfnamefont {K.~S.}\ \bibnamefont
  {Novoselov}}, \bibinfo {author} {\bibfnamefont {A.~K.}\ \bibnamefont {Geim}},
  \bibinfo {author} {\bibfnamefont {S.~V.}\ \bibnamefont {Morozov}}, \bibinfo
  {author} {\bibfnamefont {D.}~\bibnamefont {Jiang}}, \bibinfo {author}
  {\bibfnamefont {M.~I.}\ \bibnamefont {Katsnelson}}, \bibinfo {author}
  {\bibfnamefont {I.~V.}\ \bibnamefont {Grigorieva}}, \bibinfo {author}
  {\bibfnamefont {S.~V.}\ \bibnamefont {Dubonos}}, \ and\ \bibinfo {author}
  {\bibfnamefont {A.~A.}\ \bibnamefont {Firsov}},\ }\href
  {http://dx.doi.org/10.1038/nature04233} {\bibfield  {journal} {\bibinfo
  {journal} {Nature}\ }\textbf {\bibinfo {volume} {438}},\ \bibinfo {pages}
  {197 } (\bibinfo {year} {2005})}\BibitemShut {NoStop}%
\bibitem [{\citenamefont {Miro}\ \emph {et~al.}(2014)\citenamefont {Miro},
  \citenamefont {Audiffred},\ and\ \citenamefont {Heine}}]{atlas2D}%
  \BibitemOpen
  \bibfield  {author} {\bibinfo {author} {\bibfnamefont {P.}~\bibnamefont
  {Miro}}, \bibinfo {author} {\bibfnamefont {M.}~\bibnamefont {Audiffred}}, \
  and\ \bibinfo {author} {\bibfnamefont {T.}~\bibnamefont {Heine}},\ }\href
  {\doibase 10.1039/C4CS00102H} {\bibfield  {journal} {\bibinfo  {journal}
  {Chem. Soc. Rev.}\ }\textbf {\bibinfo {volume} {43}},\ \bibinfo {pages}
  {6537} (\bibinfo {year} {2014})}\BibitemShut {NoStop}%
\bibitem [{\citenamefont {Mak}\ and\ \citenamefont
  {Shan}(2016)}]{TMDCs_photonics}%
  \BibitemOpen
  \bibfield  {author} {\bibinfo {author} {\bibfnamefont {K.~F.}\ \bibnamefont
  {Mak}}\ and\ \bibinfo {author} {\bibfnamefont {J.}~\bibnamefont {Shan}},\
  }\href {http://dx.doi.org/10.1038/nphoton.2015.282} {\bibfield  {journal}
  {\bibinfo  {journal} {Nat. Photon}\ }\textbf {\bibinfo {volume} {10}},\
  \bibinfo {pages} {216 } (\bibinfo {year} {2016})}\BibitemShut {NoStop}%
\bibitem [{\citenamefont {Geim}\ and\ \citenamefont {Novoselov}(2007)}]{Geim}%
  \BibitemOpen
  \bibfield  {author} {\bibinfo {author} {\bibfnamefont {A.~K.}\ \bibnamefont
  {Geim}}\ and\ \bibinfo {author} {\bibfnamefont {K.~S.}\ \bibnamefont
  {Novoselov}},\ }\href {http://dx.doi.org/10.1038/nmat1849} {\bibfield
  {journal} {\bibinfo  {journal} {Nat. Mater.}\ }\textbf {\bibinfo {volume}
  {6}},\ \bibinfo {pages} {183} (\bibinfo {year} {2007})}\BibitemShut {NoStop}%
\bibitem [{\citenamefont {Li}\ and\ \citenamefont {Zhu}(2015)}]{Li201533}%
  \BibitemOpen
  \bibfield  {author} {\bibinfo {author} {\bibfnamefont {X.}~\bibnamefont
  {Li}}\ and\ \bibinfo {author} {\bibfnamefont {H.}~\bibnamefont {Zhu}},\
  }\href {\doibase http://dx.doi.org/10.1016/j.jmat.2015.03.003} {\bibfield
  {journal} {\bibinfo  {journal} {J. Materiomics}\ }\textbf {\bibinfo {volume}
  {1}},\ \bibinfo {pages} {33 } (\bibinfo {year} {2015})}\BibitemShut {NoStop}%
\bibitem [{\citenamefont {Wang}\ \emph {et~al.}(2012)\citenamefont {Wang},
  \citenamefont {Kalantar-Zadeh}, \citenamefont {Kis}, \citenamefont
  {Coleman},\ and\ \citenamefont {Strano}}]{GeneralFormula}%
  \BibitemOpen
  \bibfield  {author} {\bibinfo {author} {\bibfnamefont {Q.~H.}\ \bibnamefont
  {Wang}}, \bibinfo {author} {\bibfnamefont {K.}~\bibnamefont
  {Kalantar-Zadeh}}, \bibinfo {author} {\bibfnamefont {A.}~\bibnamefont {Kis}},
  \bibinfo {author} {\bibfnamefont {J.~N.}\ \bibnamefont {Coleman}}, \ and\
  \bibinfo {author} {\bibfnamefont {M.~S.}\ \bibnamefont {Strano}},\ }\href
  {http://dx.doi.org/10.1038/nnano.2012.193} {\bibfield  {journal} {\bibinfo
  {journal} {Nat. Nanotechnol.}\ }\textbf {\bibinfo {volume} {7}},\ \bibinfo
  {pages} {699} (\bibinfo {year} {2012})}\BibitemShut {NoStop}%
\bibitem [{\citenamefont {Mir\'{o}}\ \emph {et~al.}(2014)\citenamefont
  {Mir\'{o}}, \citenamefont {Han}, \citenamefont {Cheon},\ and\ \citenamefont
  {Heine}}]{NanoflakesMiro}%
  \BibitemOpen
  \bibfield  {author} {\bibinfo {author} {\bibfnamefont {P.}~\bibnamefont
  {Mir\'{o}}}, \bibinfo {author} {\bibfnamefont {J.~H.}\ \bibnamefont {Han}},
  \bibinfo {author} {\bibfnamefont {J.}~\bibnamefont {Cheon}}, \ and\ \bibinfo
  {author} {\bibfnamefont {T.}~\bibnamefont {Heine}},\ }\href {\doibase
  10.1002/anie.201404704} {\bibfield  {journal} {\bibinfo  {journal} {Angew.
  Chem. Int. Ed.}\ }\textbf {\bibinfo {volume} {53}},\ \bibinfo {pages} {12624}
  (\bibinfo {year} {2014})}\BibitemShut {NoStop}%
\bibitem [{\citenamefont {{Splendiani}}\ \emph {et~al.}(2010)\citenamefont
  {{Splendiani}}, \citenamefont {{Sun}}, \citenamefont {{Zhang}}, \citenamefont
  {{Li}}, \citenamefont {{Kim}}, \citenamefont {{Chim}}, \citenamefont
  {{Galli}},\ and\ \citenamefont {{Wang}}}]{2010Nano10}%
  \BibitemOpen
  \bibfield  {author} {\bibinfo {author} {\bibfnamefont {A.}~\bibnamefont
  {{Splendiani}}}, \bibinfo {author} {\bibfnamefont {L.}~\bibnamefont {{Sun}}},
  \bibinfo {author} {\bibfnamefont {Y.}~\bibnamefont {{Zhang}}}, \bibinfo
  {author} {\bibfnamefont {T.}~\bibnamefont {{Li}}}, \bibinfo {author}
  {\bibfnamefont {J.}~\bibnamefont {{Kim}}}, \bibinfo {author} {\bibfnamefont
  {C.-Y.}\ \bibnamefont {{Chim}}}, \bibinfo {author} {\bibfnamefont
  {G.}~\bibnamefont {{Galli}}}, \ and\ \bibinfo {author} {\bibfnamefont
  {F.}~\bibnamefont {{Wang}}},\ }\href {\doibase 10.1021/nl903868w} {\bibfield
  {journal} {\bibinfo  {journal} {Nano Lett.}\ }\textbf {\bibinfo {volume}
  {10}},\ \bibinfo {pages} {1271} (\bibinfo {year} {2010})}\BibitemShut
  {NoStop}%
\bibitem [{\citenamefont {Mak}\ \emph {et~al.}(2010)\citenamefont {Mak},
  \citenamefont {Lee}, \citenamefont {Hone}, \citenamefont {Shan},\ and\
  \citenamefont {Heinz}}]{PhysRevLett}%
  \BibitemOpen
  \bibfield  {author} {\bibinfo {author} {\bibfnamefont {K.~F.}\ \bibnamefont
  {Mak}}, \bibinfo {author} {\bibfnamefont {C.}~\bibnamefont {Lee}}, \bibinfo
  {author} {\bibfnamefont {J.}~\bibnamefont {Hone}}, \bibinfo {author}
  {\bibfnamefont {J.}~\bibnamefont {Shan}}, \ and\ \bibinfo {author}
  {\bibfnamefont {T.~F.}\ \bibnamefont {Heinz}},\ }\href {\doibase
  10.1103/PhysRevLett.105.136805} {\bibfield  {journal} {\bibinfo  {journal}
  {Phys. Rev. Lett.}\ }\textbf {\bibinfo {volume} {105}},\ \bibinfo {pages}
  {136805} (\bibinfo {year} {2010})}\BibitemShut {NoStop}%
\bibitem [{\citenamefont {Kadantsev}\ and\ \citenamefont
  {Hawrylak}(2012)}]{Kadantsev2012909}%
  \BibitemOpen
  \bibfield  {author} {\bibinfo {author} {\bibfnamefont {E.~S.}\ \bibnamefont
  {Kadantsev}}\ and\ \bibinfo {author} {\bibfnamefont {P.}~\bibnamefont
  {Hawrylak}},\ }\href {\doibase http://dx.doi.org/10.1016/j.ssc.2012.02.005}
  {\bibfield  {journal} {\bibinfo  {journal} {Solid State Commun.}\ }\textbf
  {\bibinfo {volume} {152}},\ \bibinfo {pages} {909 } (\bibinfo {year}
  {2012})}\BibitemShut {NoStop}%
\bibitem [{\citenamefont {Xiao}\ \emph {et~al.}(2012)\citenamefont {Xiao},
  \citenamefont {Liu}, \citenamefont {Feng}, \citenamefont {Xu},\ and\
  \citenamefont {Yao}}]{XLF2012}%
  \BibitemOpen
  \bibfield  {author} {\bibinfo {author} {\bibfnamefont {D.}~\bibnamefont
  {Xiao}}, \bibinfo {author} {\bibfnamefont {G.-B.}\ \bibnamefont {Liu}},
  \bibinfo {author} {\bibfnamefont {W.}~\bibnamefont {Feng}}, \bibinfo {author}
  {\bibfnamefont {X.}~\bibnamefont {Xu}}, \ and\ \bibinfo {author}
  {\bibfnamefont {W.}~\bibnamefont {Yao}},\ }\href {\doibase
  10.1103/PhysRevLett.108.196802} {\bibfield  {journal} {\bibinfo  {journal}
  {Phys. Rev. Lett.}\ }\textbf {\bibinfo {volume} {108}},\ \bibinfo {pages}
  {196802} (\bibinfo {year} {2012})}\BibitemShut {NoStop}%
\bibitem [{\citenamefont {Wendumu}\ \emph {et~al.}(2014)\citenamefont
  {Wendumu}, \citenamefont {Seifert}, \citenamefont {Lorenz}, \citenamefont
  {Joswig},\ and\ \citenamefont {Enyashin}}]{Wendumu2014}%
  \BibitemOpen
  \bibfield  {author} {\bibinfo {author} {\bibfnamefont {T.~B.}\ \bibnamefont
  {Wendumu}}, \bibinfo {author} {\bibfnamefont {G.}~\bibnamefont {Seifert}},
  \bibinfo {author} {\bibfnamefont {T.}~\bibnamefont {Lorenz}}, \bibinfo
  {author} {\bibfnamefont {J.-O.}\ \bibnamefont {Joswig}}, \ and\ \bibinfo
  {author} {\bibfnamefont {A.}~\bibnamefont {Enyashin}},\ }\href {\doibase
  10.1021/jz501604j} {\bibfield  {journal} {\bibinfo  {journal} {J. Phys. Chem.
  Lett.}\ }\textbf {\bibinfo {volume} {5}},\ \bibinfo {pages} {3636} (\bibinfo
  {year} {2014})}\BibitemShut {NoStop}%
\bibitem [{\citenamefont {Pan}\ and\ \citenamefont {Zhang}(2012)}]{PZ2012}%
  \BibitemOpen
  \bibfield  {author} {\bibinfo {author} {\bibfnamefont {H.}~\bibnamefont
  {Pan}}\ and\ \bibinfo {author} {\bibfnamefont {Y.-W.}\ \bibnamefont
  {Zhang}},\ }\href {\doibase 10.1039/C2JM15906F} {\bibfield  {journal}
  {\bibinfo  {journal} {J. Mater. Chem.}\ }\textbf {\bibinfo {volume} {22}},\
  \bibinfo {pages} {7280} (\bibinfo {year} {2012})}\BibitemShut {NoStop}%
\bibitem [{\citenamefont {Nguyen}\ \emph {et~al.}(2016)\citenamefont {Nguyen},
  \citenamefont {Sohn}, \citenamefont {Oh}, \citenamefont {Jang},\ and\
  \citenamefont {Kim}}]{Nguyen2016}%
  \BibitemOpen
  \bibfield  {author} {\bibinfo {author} {\bibfnamefont {T.~P.}\ \bibnamefont
  {Nguyen}}, \bibinfo {author} {\bibfnamefont {W.}~\bibnamefont {Sohn}},
  \bibinfo {author} {\bibfnamefont {J.~H.}\ \bibnamefont {Oh}}, \bibinfo
  {author} {\bibfnamefont {H.~W.}\ \bibnamefont {Jang}}, \ and\ \bibinfo
  {author} {\bibfnamefont {S.~Y.}\ \bibnamefont {Kim}},\ }\href {\doibase
  10.1021/acs.jpcc.6b01838} {\bibfield  {journal} {\bibinfo  {journal} {J.
  Phys. Chem. C}\ }\textbf {\bibinfo {volume} {120}},\ \bibinfo {pages} {10078}
  (\bibinfo {year} {2016})}\BibitemShut {NoStop}%
\bibitem [{gau()}]{gaussiao09}%
  \BibitemOpen
  \href@noop {} {}\bibinfo {note}
  {\textsc{gaussian09}\textsuperscript{\textregistered} version 3.5a, Gaussian,
  Inc., Wallingford CT, 2013}\BibitemShut {NoStop}%
\bibitem [{\citenamefont {Bertram}\ \emph {et~al.}(2006)\citenamefont
  {Bertram}, \citenamefont {Cordes}, \citenamefont {Kim}, \citenamefont
  {Ganteför}, \citenamefont {Gemming},\ and\ \citenamefont
  {Seifert}}]{Bertram2006}%
  \BibitemOpen
  \bibfield  {author} {\bibinfo {author} {\bibfnamefont {N.}~\bibnamefont
  {Bertram}}, \bibinfo {author} {\bibfnamefont {J.}~\bibnamefont {Cordes}},
  \bibinfo {author} {\bibfnamefont {Y.}~\bibnamefont {Kim}}, \bibinfo {author}
  {\bibfnamefont {G.}~\bibnamefont {Ganteför}}, \bibinfo {author}
  {\bibfnamefont {S.}~\bibnamefont {Gemming}}, \ and\ \bibinfo {author}
  {\bibfnamefont {G.}~\bibnamefont {Seifert}},\ }\href {\doibase
  http://dx.doi.org/10.1016/j.cplett.2005.10.046} {\bibfield  {journal}
  {\bibinfo  {journal} {Chem. Phys. Lett.}\ }\textbf {\bibinfo {volume}
  {418}},\ \bibinfo {pages} {36} (\bibinfo {year} {2006})}\BibitemShut
  {NoStop}%
\bibitem [{\citenamefont {Lauritsen}\ \emph {et~al.}(2007)\citenamefont
  {Lauritsen}, \citenamefont {Kibsgaard}, \citenamefont {Helveg}, \citenamefont
  {Topsoe}, \citenamefont {Clausen}, \citenamefont {Laegsgaard},\ and\
  \citenamefont {Besenbacher}}]{Lauritsen2007}%
  \BibitemOpen
  \bibfield  {author} {\bibinfo {author} {\bibfnamefont {J.~V.}\ \bibnamefont
  {Lauritsen}}, \bibinfo {author} {\bibfnamefont {J.}~\bibnamefont
  {Kibsgaard}}, \bibinfo {author} {\bibfnamefont {S.}~\bibnamefont {Helveg}},
  \bibinfo {author} {\bibfnamefont {H.}~\bibnamefont {Topsoe}}, \bibinfo
  {author} {\bibfnamefont {B.~S.}\ \bibnamefont {Clausen}}, \bibinfo {author}
  {\bibfnamefont {E.}~\bibnamefont {Laegsgaard}}, \ and\ \bibinfo {author}
  {\bibfnamefont {F.}~\bibnamefont {Besenbacher}},\ }\href
  {http://dx.doi.org/10.1038/nnano.2006.171} {\bibfield  {journal} {\bibinfo
  {journal} {Nat. Nano}\ }\textbf {\bibinfo {volume} {2}},\ \bibinfo {pages}
  {53} (\bibinfo {year} {2007})}\BibitemShut {NoStop}%
\bibitem [{\citenamefont {Helveg}\ \emph {et~al.}(2000)\citenamefont {Helveg},
  \citenamefont {Lauritsen}, \citenamefont {L\ae{}gsgaard}, \citenamefont
  {Stensgaard}, \citenamefont {N\o{}rskov}, \citenamefont {Clausen},
  \citenamefont {Tops\o{}e},\ and\ \citenamefont {Besenbacher}}]{Halveg2000}%
  \BibitemOpen
  \bibfield  {author} {\bibinfo {author} {\bibfnamefont {S.}~\bibnamefont
  {Helveg}}, \bibinfo {author} {\bibfnamefont {J.~V.}\ \bibnamefont
  {Lauritsen}}, \bibinfo {author} {\bibfnamefont {E.}~\bibnamefont
  {L\ae{}gsgaard}}, \bibinfo {author} {\bibfnamefont {I.}~\bibnamefont
  {Stensgaard}}, \bibinfo {author} {\bibfnamefont {J.~K.}\ \bibnamefont
  {N\o{}rskov}}, \bibinfo {author} {\bibfnamefont {B.~S.}\ \bibnamefont
  {Clausen}}, \bibinfo {author} {\bibfnamefont {H.}~\bibnamefont {Tops\o{}e}},
  \ and\ \bibinfo {author} {\bibfnamefont {F.}~\bibnamefont {Besenbacher}},\
  }\href {\doibase 10.1103/PhysRevLett.84.951} {\bibfield  {journal} {\bibinfo
  {journal} {Phys. Rev. Lett.}\ }\textbf {\bibinfo {volume} {84}},\ \bibinfo
  {pages} {951} (\bibinfo {year} {2000})}\BibitemShut {NoStop}%
\bibitem [{\citenamefont {Seifert}\ \emph {et~al.}(2006)\citenamefont
  {Seifert}, \citenamefont {Tamuliene},\ and\ \citenamefont
  {Gemming}}]{Seifert2006}%
  \BibitemOpen
  \bibfield  {author} {\bibinfo {author} {\bibfnamefont {G.}~\bibnamefont
  {Seifert}}, \bibinfo {author} {\bibfnamefont {J.}~\bibnamefont {Tamuliene}},
  \ and\ \bibinfo {author} {\bibfnamefont {S.}~\bibnamefont {Gemming}},\ }\href
  {\doibase http://dx.doi.org/10.1016/j.commatsci.2004.08.014} {\bibfield
  {journal} {\bibinfo  {journal} {Comput. Mater. Sci}\ }\textbf {\bibinfo
  {volume} {35}},\ \bibinfo {pages} {316} (\bibinfo {year} {2006})}\BibitemShut
  {NoStop}%
\bibitem [{\citenamefont {Dickinson}\ and\ \citenamefont
  {Pauling}(1923)}]{Dickinson1923}%
  \BibitemOpen
  \bibfield  {author} {\bibinfo {author} {\bibfnamefont {R.~G.}\ \bibnamefont
  {Dickinson}}\ and\ \bibinfo {author} {\bibfnamefont {L.}~\bibnamefont
  {Pauling}},\ }\href {\doibase 10.1021/ja01659a020} {\bibfield  {journal}
  {\bibinfo  {journal} {J. Am. Chem. Soc.}\ }\textbf {\bibinfo {volume} {45}},\
  \bibinfo {pages} {1466} (\bibinfo {year} {1923})}\BibitemShut {NoStop}%
\bibitem [{\citenamefont {Becke}(1988)}]{Becke88}%
  \BibitemOpen
  \bibfield  {author} {\bibinfo {author} {\bibfnamefont {A.~D.}\ \bibnamefont
  {Becke}},\ }\href {\doibase 10.1103/PhysRevA.38.3098} {\bibfield  {journal}
  {\bibinfo  {journal} {Phys. Rev. A}\ }\textbf {\bibinfo {volume} {38}},\
  \bibinfo {pages} {3098} (\bibinfo {year} {1988})}\BibitemShut {NoStop}%
\bibitem [{\citenamefont {Lee}\ \emph {et~al.}(1988)\citenamefont {Lee},
  \citenamefont {Yang},\ and\ \citenamefont {Parr}}]{ECLYP}%
  \BibitemOpen
  \bibfield  {author} {\bibinfo {author} {\bibfnamefont {C.}~\bibnamefont
  {Lee}}, \bibinfo {author} {\bibfnamefont {W.}~\bibnamefont {Yang}}, \ and\
  \bibinfo {author} {\bibfnamefont {R.~G.}\ \bibnamefont {Parr}},\ }\href
  {\doibase 10.1103/PhysRevB.37.785} {\bibfield  {journal} {\bibinfo  {journal}
  {Phys. Rev. B}\ }\textbf {\bibinfo {volume} {37}},\ \bibinfo {pages} {785}
  (\bibinfo {year} {1988})}\BibitemShut {NoStop}%
\bibitem [{\citenamefont {Stephens}\ \emph {et~al.}(1994)\citenamefont
  {Stephens}, \citenamefont {Devlin}, \citenamefont {Chabalowski},\ and\
  \citenamefont {Frisch}}]{B3LYP}%
  \BibitemOpen
  \bibfield  {author} {\bibinfo {author} {\bibfnamefont {P.~J.}\ \bibnamefont
  {Stephens}}, \bibinfo {author} {\bibfnamefont {F.~J.}\ \bibnamefont
  {Devlin}}, \bibinfo {author} {\bibfnamefont {C.~F.}\ \bibnamefont
  {Chabalowski}}, \ and\ \bibinfo {author} {\bibfnamefont {M.~J.}\ \bibnamefont
  {Frisch}},\ }\href {\doibase 10.1021/j100096a001} {\bibfield  {journal}
  {\bibinfo  {journal} {J. Phys. Chem.}\ }\textbf {\bibinfo {volume} {98}},\
  \bibinfo {pages} {11623} (\bibinfo {year} {1994})}\BibitemShut {NoStop}%
\bibitem [{\citenamefont {Becke}(1993{\natexlab{a}})}]{Functional}%
  \BibitemOpen
  \bibfield  {author} {\bibinfo {author} {\bibfnamefont {A.~D.}\ \bibnamefont
  {Becke}},\ }\href {\doibase http://dx.doi.org/10.1063/1.464304} {\bibfield
  {journal} {\bibinfo  {journal} {J. Chem. Phys.}\ }\textbf {\bibinfo {volume}
  {98}},\ \bibinfo {pages} {1372} (\bibinfo {year}
  {1993}{\natexlab{a}})}\BibitemShut {NoStop}%
\bibitem [{\citenamefont {Adamo}\ and\ \citenamefont {Barone}(1999)}]{PBE1PBE}%
  \BibitemOpen
  \bibfield  {author} {\bibinfo {author} {\bibfnamefont {C.}~\bibnamefont
  {Adamo}}\ and\ \bibinfo {author} {\bibfnamefont {V.}~\bibnamefont {Barone}},\
  }\href {\doibase 10.1063/1.478522} {\bibfield  {journal} {\bibinfo  {journal}
  {J. Chem. Phys.}\ }\textbf {\bibinfo {volume} {110}},\ \bibinfo {pages}
  {6158} (\bibinfo {year} {1999})}\BibitemShut {NoStop}%
\bibitem [{\citenamefont {Zhao}\ \emph {et~al.}(2006)\citenamefont {Zhao},
  \citenamefont {Schultz},\ and\ \citenamefont {Truhlar}}]{M052XFunctional}%
  \BibitemOpen
  \bibfield  {author} {\bibinfo {author} {\bibfnamefont {Y.}~\bibnamefont
  {Zhao}}, \bibinfo {author} {\bibfnamefont {N.~E.}\ \bibnamefont {Schultz}}, \
  and\ \bibinfo {author} {\bibfnamefont {D.~G.}\ \bibnamefont {Truhlar}},\
  }\href {\doibase 10.1021/ct0502763} {\bibfield  {journal} {\bibinfo
  {journal} {J. Chem. Theory Comput.}\ }\textbf {\bibinfo {volume} {2}},\
  \bibinfo {pages} {364} (\bibinfo {year} {2006})}\BibitemShut {NoStop}%
\bibitem [{\citenamefont {Gan}\ \emph {et~al.}(2015)\citenamefont {Gan},
  \citenamefont {Liu}, \citenamefont {Wu}, \citenamefont {Hao}, \citenamefont
  {Shan}, \citenamefont {Wu},\ and\ \citenamefont {Chu}}]{Gan2015}%
  \BibitemOpen
  \bibfield  {author} {\bibinfo {author} {\bibfnamefont {Z.~X.}\ \bibnamefont
  {Gan}}, \bibinfo {author} {\bibfnamefont {L.~Z.}\ \bibnamefont {Liu}},
  \bibinfo {author} {\bibfnamefont {H.~Y.}\ \bibnamefont {Wu}}, \bibinfo
  {author} {\bibfnamefont {Y.~L.}\ \bibnamefont {Hao}}, \bibinfo {author}
  {\bibfnamefont {Y.}~\bibnamefont {Shan}}, \bibinfo {author} {\bibfnamefont
  {X.~L.}\ \bibnamefont {Wu}}, \ and\ \bibinfo {author} {\bibfnamefont {P.~K.}\
  \bibnamefont {Chu}},\ }\href {\doibase 10.1063/1.4922551} {\bibfield
  {journal} {\bibinfo  {journal} {Appl. Phys. Lett.}\ }\textbf {\bibinfo
  {volume} {106}},\ \bibinfo {pages} {233113} (\bibinfo {year}
  {2015})}\BibitemShut {NoStop}%
\bibitem [{\citenamefont {Cramer}\ and\ \citenamefont
  {Truhlar}(2009)}]{M052XReview}%
  \BibitemOpen
  \bibfield  {author} {\bibinfo {author} {\bibfnamefont {C.~J.}\ \bibnamefont
  {Cramer}}\ and\ \bibinfo {author} {\bibfnamefont {D.~G.}\ \bibnamefont
  {Truhlar}},\ }\href {\doibase 10.1039/B907148B} {\bibfield  {journal}
  {\bibinfo  {journal} {Phys. Chem. Chem. Phys.}\ }\textbf {\bibinfo {volume}
  {11}},\ \bibinfo {pages} {10757} (\bibinfo {year} {2009})}\BibitemShut
  {NoStop}%
\bibitem [{\citenamefont {Slater}(1974)}]{Slater}%
  \BibitemOpen
  \bibfield  {author} {\bibinfo {author} {\bibfnamefont {J.~C.}\ \bibnamefont
  {Slater}},\ }\href@noop {} {\emph {\bibinfo {title} {Quantum theory of
  molecules and solids. vol. 4. The self-consistent field for molecules and
  solids}}}\ (\bibinfo  {publisher} {New York NY : McGraw-Hill},\ \bibinfo
  {year} {1974})\BibitemShut {NoStop}%
\bibitem [{\citenamefont {Chiodo}\ \emph {et~al.}(2006)\citenamefont {Chiodo},
  \citenamefont {Russo},\ and\ \citenamefont {Sicilia}}]{LANL2DZ}%
  \BibitemOpen
  \bibfield  {author} {\bibinfo {author} {\bibfnamefont {S.}~\bibnamefont
  {Chiodo}}, \bibinfo {author} {\bibfnamefont {N.}~\bibnamefont {Russo}}, \
  and\ \bibinfo {author} {\bibfnamefont {E.}~\bibnamefont {Sicilia}},\ }\href
  {\doibase 10.1063/1.2345197} {\bibfield  {journal} {\bibinfo  {journal} {J.
  Chem. Phys.}\ }\textbf {\bibinfo {volume} {125}},\ \bibinfo {pages} {104107}
  (\bibinfo {year} {2006})}\BibitemShut {NoStop}%
\bibitem [{\citenamefont {Hay}\ and\ \citenamefont
  {Wadt}(1985{\natexlab{a}})}]{HAY85}%
  \BibitemOpen
  \bibfield  {author} {\bibinfo {author} {\bibfnamefont {P.~J.}\ \bibnamefont
  {Hay}}\ and\ \bibinfo {author} {\bibfnamefont {W.~R.}\ \bibnamefont {Wadt}},\
  }\href {\doibase http://dx.doi.org/10.1063/1.448799} {\bibfield  {journal}
  {\bibinfo  {journal} {J. Chem. Phys}\ }\textbf {\bibinfo {volume} {82}},\
  \bibinfo {pages} {270} (\bibinfo {year} {1985}{\natexlab{a}})}\BibitemShut
  {NoStop}%
\bibitem [{\citenamefont {Hay}\ and\ \citenamefont
  {Wadt}(1985{\natexlab{b}})}]{HAY85a}%
  \BibitemOpen
  \bibfield  {author} {\bibinfo {author} {\bibfnamefont {P.~J.}\ \bibnamefont
  {Hay}}\ and\ \bibinfo {author} {\bibfnamefont {W.~R.}\ \bibnamefont {Wadt}},\
  }\href {\doibase http://dx.doi.org/10.1063/1.448975} {\bibfield  {journal}
  {\bibinfo  {journal} {J. Chem. Phys.}\ }\textbf {\bibinfo {volume} {82}},\
  \bibinfo {pages} {299} (\bibinfo {year} {1985}{\natexlab{b}})}\BibitemShut
  {NoStop}%
\bibitem [{\citenamefont {Wadt}\ and\ \citenamefont {Hay}(1985)}]{Wadt85}%
  \BibitemOpen
  \bibfield  {author} {\bibinfo {author} {\bibfnamefont {W.~R.}\ \bibnamefont
  {Wadt}}\ and\ \bibinfo {author} {\bibfnamefont {P.~J.}\ \bibnamefont {Hay}},\
  }\href {\doibase http://dx.doi.org/10.1063/1.448800} {\bibfield  {journal}
  {\bibinfo  {journal} {J. Chem. Phys.}\ }\textbf {\bibinfo {volume} {82}},\
  \bibinfo {pages} {284} (\bibinfo {year} {1985})}\BibitemShut {NoStop}%
\bibitem [{\citenamefont {Li}\ and\ \citenamefont
  {Frisch}(2006)}]{BernyGEDIIS}%
  \BibitemOpen
  \bibfield  {author} {\bibinfo {author} {\bibfnamefont {X.}~\bibnamefont
  {Li}}\ and\ \bibinfo {author} {\bibfnamefont {M.~J.}\ \bibnamefont
  {Frisch}},\ }\href {\doibase 10.1021/ct050275a} {\bibfield  {journal}
  {\bibinfo  {journal} {J. Chem. Theory Comput.}\ }\textbf {\bibinfo {volume}
  {2}},\ \bibinfo {pages} {835} (\bibinfo {year} {2006})}\BibitemShut {NoStop}%
\bibitem [{Avo()}]{Avogadro}%
  \BibitemOpen
  \href@noop {} {}\bibinfo {note} {Avogadro: an open-source molecular builder
  and visualization tool. Version 1.1.1
  http://avogadro.openmolecules.net/}\BibitemShut {NoStop}%
\bibitem [{\citenamefont {Hanwell}\ \emph {et~al.}(2012)\citenamefont
  {Hanwell}, \citenamefont {Curtis}, \citenamefont {Lonie}, \citenamefont
  {Vandermeersch}, \citenamefont {Zurek},\ and\ \citenamefont
  {Hutchison}}]{Hanwell2012}%
  \BibitemOpen
  \bibfield  {author} {\bibinfo {author} {\bibfnamefont {M.~D.}\ \bibnamefont
  {Hanwell}}, \bibinfo {author} {\bibfnamefont {D.~E.}\ \bibnamefont {Curtis}},
  \bibinfo {author} {\bibfnamefont {D.~C.}\ \bibnamefont {Lonie}}, \bibinfo
  {author} {\bibfnamefont {T.}~\bibnamefont {Vandermeersch}}, \bibinfo {author}
  {\bibfnamefont {E.}~\bibnamefont {Zurek}}, \ and\ \bibinfo {author}
  {\bibfnamefont {G.~R.}\ \bibnamefont {Hutchison}},\ }\href {\doibase
  10.1186/1758-2946-4-17} {\bibfield  {journal} {\bibinfo  {journal} {J.
  Cheminform}\ }\textbf {\bibinfo {volume} {4}},\ \bibinfo {pages} {1}
  (\bibinfo {year} {2012})}\BibitemShut {NoStop}%
\bibitem [{\citenamefont {Topsoe}\ and\ \citenamefont
  {Topsoe}(1993)}]{TOPSOE1993}%
  \BibitemOpen
  \bibfield  {author} {\bibinfo {author} {\bibfnamefont {N.}~\bibnamefont
  {Topsoe}}\ and\ \bibinfo {author} {\bibfnamefont {H.}~\bibnamefont
  {Topsoe}},\ }\href {\doibase http://dx.doi.org/10.1006/jcat.1993.1056}
  {\bibfield  {journal} {\bibinfo  {journal} {J. Catal.}\ }\textbf {\bibinfo
  {volume} {139}},\ \bibinfo {pages} {641} (\bibinfo {year}
  {1993})}\BibitemShut {NoStop}%
\bibitem [{\citenamefont {Loh}\ \emph {et~al.}(2015)\citenamefont {Loh},
  \citenamefont {Pandey}, \citenamefont {Yap},\ and\ \citenamefont
  {Karna}}]{Passivation}%
  \BibitemOpen
  \bibfield  {author} {\bibinfo {author} {\bibfnamefont {G.~C.}\ \bibnamefont
  {Loh}}, \bibinfo {author} {\bibfnamefont {R.}~\bibnamefont {Pandey}},
  \bibinfo {author} {\bibfnamefont {Y.~K.}\ \bibnamefont {Yap}}, \ and\
  \bibinfo {author} {\bibfnamefont {S.~P.}\ \bibnamefont {Karna}},\ }\href
  {\doibase 10.1021/jp510598x} {\bibfield  {journal} {\bibinfo  {journal} {J.
  Phys. Chem. C}\ }\textbf {\bibinfo {volume} {119}},\ \bibinfo {pages} {1565}
  (\bibinfo {year} {2015})}\BibitemShut {NoStop}%
\bibitem [{\citenamefont {Carruth}\ and\ \citenamefont
  {Eugene}(2002)}]{Gorton}%
  \BibitemOpen
  \bibfield  {author} {\bibinfo {author} {\bibfnamefont {G.}~\bibnamefont
  {Carruth}}\ and\ \bibinfo {author} {\bibfnamefont {E.}~\bibnamefont
  {Eugene}},\ }\href@noop {} {\emph {\bibinfo {title} {Bond energies}}}\
  (\bibinfo  {publisher} {Vol. 1, Tennessee: Southwestern},\ \bibinfo {year}
  {2002})\BibitemShut {NoStop}%
\bibitem [{\citenamefont {Perdew}(1986)}]{BP86}%
  \BibitemOpen
  \bibfield  {author} {\bibinfo {author} {\bibfnamefont {J.~P.}\ \bibnamefont
  {Perdew}},\ }\href {\doibase 10.1103/PhysRevB.33.8822} {\bibfield  {journal}
  {\bibinfo  {journal} {Phys. Rev. B}\ }\textbf {\bibinfo {volume} {33}},\
  \bibinfo {pages} {8822} (\bibinfo {year} {1986})}\BibitemShut {NoStop}%
\bibitem [{\citenamefont {Heyd}\ and\ \citenamefont
  {Scuseria}(2004)}]{HSEH1PBE_1}%
  \BibitemOpen
  \bibfield  {author} {\bibinfo {author} {\bibfnamefont {J.}~\bibnamefont
  {Heyd}}\ and\ \bibinfo {author} {\bibfnamefont {G.~E.}\ \bibnamefont
  {Scuseria}},\ }\href {\doibase 10.1063/1.1760074} {\bibfield  {journal}
  {\bibinfo  {journal} {J. Chem. Phys.}\ }\textbf {\bibinfo {volume} {121}},\
  \bibinfo {pages} {1187} (\bibinfo {year} {2004})}\BibitemShut {NoStop}%
\bibitem [{\citenamefont {Krukau}\ \emph {et~al.}(2006)\citenamefont {Krukau},
  \citenamefont {Vydrov}, \citenamefont {Izmaylov},\ and\ \citenamefont
  {Scuseria}}]{HSEH1PBE_2}%
  \BibitemOpen
  \bibfield  {author} {\bibinfo {author} {\bibfnamefont {A.~V.}\ \bibnamefont
  {Krukau}}, \bibinfo {author} {\bibfnamefont {O.~A.}\ \bibnamefont {Vydrov}},
  \bibinfo {author} {\bibfnamefont {A.~F.}\ \bibnamefont {Izmaylov}}, \ and\
  \bibinfo {author} {\bibfnamefont {G.~E.}\ \bibnamefont {Scuseria}},\ }\href
  {\doibase 10.1063/1.2404663} {\bibfield  {journal} {\bibinfo  {journal} {J.
  Chem. Phys.}\ }\textbf {\bibinfo {volume} {125}},\ \bibinfo {pages} {224106}
  (\bibinfo {year} {2006})}\BibitemShut {NoStop}%
\bibitem [{\citenamefont {Izmaylov}\ \emph {et~al.}(2006)\citenamefont
  {Izmaylov}, \citenamefont {Scuseria},\ and\ \citenamefont
  {Frisch}}]{HSEH1PBE_3}%
  \BibitemOpen
  \bibfield  {author} {\bibinfo {author} {\bibfnamefont {A.~F.}\ \bibnamefont
  {Izmaylov}}, \bibinfo {author} {\bibfnamefont {G.~E.}\ \bibnamefont
  {Scuseria}}, \ and\ \bibinfo {author} {\bibfnamefont {M.~J.}\ \bibnamefont
  {Frisch}},\ }\href {\doibase 10.1063/1.2347713} {\bibfield  {journal}
  {\bibinfo  {journal} {J. Chem. Phys.}\ }\textbf {\bibinfo {volume} {125}},\
  \bibinfo {pages} {104103} (\bibinfo {year} {2006})}\BibitemShut {NoStop}%
\bibitem [{\citenamefont {Henderson}\ \emph {et~al.}(2009)\citenamefont
  {Henderson}, \citenamefont {Izmaylov}, \citenamefont {Scalmani},\ and\
  \citenamefont {Scuseria}}]{HSEH1PBE_4}%
  \BibitemOpen
  \bibfield  {author} {\bibinfo {author} {\bibfnamefont {T.~M.}\ \bibnamefont
  {Henderson}}, \bibinfo {author} {\bibfnamefont {A.~F.}\ \bibnamefont
  {Izmaylov}}, \bibinfo {author} {\bibfnamefont {G.}~\bibnamefont {Scalmani}},
  \ and\ \bibinfo {author} {\bibfnamefont {G.~E.}\ \bibnamefont {Scuseria}},\
  }\href {\doibase 10.1063/1.3185673} {\bibfield  {journal} {\bibinfo
  {journal} {J. Chem. Phys.}\ }\textbf {\bibinfo {volume} {131}},\ \bibinfo
  {pages} {044108} (\bibinfo {year} {2009})}\BibitemShut {NoStop}%
\bibitem [{\citenamefont {Heyd}\ \emph {et~al.}(2006)\citenamefont {Heyd},
  \citenamefont {Scuseria},\ and\ \citenamefont {Ernzerhof}}]{HSEH1PBE_5}%
  \BibitemOpen
  \bibfield  {author} {\bibinfo {author} {\bibfnamefont {J.}~\bibnamefont
  {Heyd}}, \bibinfo {author} {\bibfnamefont {G.~E.}\ \bibnamefont {Scuseria}},
  \ and\ \bibinfo {author} {\bibfnamefont {M.}~\bibnamefont {Ernzerhof}},\
  }\href {\doibase 10.1063/1.2204597} {\bibfield  {journal} {\bibinfo
  {journal} {J. Chem. Phys.}\ }\textbf {\bibinfo {volume} {124}},\ \bibinfo
  {pages} {219906} (\bibinfo {year} {2006})}\BibitemShut {NoStop}%
\bibitem [{\citenamefont {Heyd}\ \emph {et~al.}(2005)\citenamefont {Heyd},
  \citenamefont {Peralta}, \citenamefont {Scuseria},\ and\ \citenamefont
  {Martin}}]{HSEH1PBE_6}%
  \BibitemOpen
  \bibfield  {author} {\bibinfo {author} {\bibfnamefont {J.}~\bibnamefont
  {Heyd}}, \bibinfo {author} {\bibfnamefont {J.~E.}\ \bibnamefont {Peralta}},
  \bibinfo {author} {\bibfnamefont {G.~E.}\ \bibnamefont {Scuseria}}, \ and\
  \bibinfo {author} {\bibfnamefont {R.~L.}\ \bibnamefont {Martin}},\ }\href
  {\doibase 10.1063/1.2085170} {\bibfield  {journal} {\bibinfo  {journal} {J.
  Chem. Phys.}\ }\textbf {\bibinfo {volume} {123}},\ \bibinfo {pages} {174101}
  (\bibinfo {year} {2005})}\BibitemShut {NoStop}%
\bibitem [{\citenamefont {Becke}(1993{\natexlab{b}})}]{B3PW91}%
  \BibitemOpen
  \bibfield  {author} {\bibinfo {author} {\bibfnamefont {A.~D.}\ \bibnamefont
  {Becke}},\ }\href {\doibase 10.1063/1.464913} {\bibfield  {journal} {\bibinfo
   {journal} {J. Chem. Phys.}\ }\textbf {\bibinfo {volume} {98}},\ \bibinfo
  {pages} {5648} (\bibinfo {year} {1993}{\natexlab{b}})}\BibitemShut {NoStop}%
\bibitem [{\citenamefont {Ernzerhof}\ and\ \citenamefont
  {Perdew}(1998)}]{PBEh1PBE}%
  \BibitemOpen
  \bibfield  {author} {\bibinfo {author} {\bibfnamefont {M.}~\bibnamefont
  {Ernzerhof}}\ and\ \bibinfo {author} {\bibfnamefont {J.~P.}\ \bibnamefont
  {Perdew}},\ }\href {\doibase 10.1063/1.476928} {\bibfield  {journal}
  {\bibinfo  {journal} {J. Chem. Phys.}\ }\textbf {\bibinfo {volume} {109}},\
  \bibinfo {pages} {3313} (\bibinfo {year} {1998})}\BibitemShut {NoStop}%
\bibitem [{\citenamefont {Zhao}\ \emph {et~al.}(2005)\citenamefont {Zhao},
  \citenamefont {Schultz},\ and\ \citenamefont {Truhlar}}]{M05}%
  \BibitemOpen
  \bibfield  {author} {\bibinfo {author} {\bibfnamefont {Y.}~\bibnamefont
  {Zhao}}, \bibinfo {author} {\bibfnamefont {N.~E.}\ \bibnamefont {Schultz}}, \
  and\ \bibinfo {author} {\bibfnamefont {D.~G.}\ \bibnamefont {Truhlar}},\
  }\href {\doibase 10.1063/1.2126975} {\bibfield  {journal} {\bibinfo
  {journal} {J. Chem. Phys.}\ }\textbf {\bibinfo {volume} {123}},\ \bibinfo
  {pages} {161103} (\bibinfo {year} {2005})}\BibitemShut {NoStop}%
\end{thebibliography}%

\end{document}